\documentclass[aps,prl,reprint,showpacs,superscriptaddress]{revtex4-1}
\usepackage{epsfig}
\usepackage[]{natbib}
\usepackage{xspace}
\usepackage[pdftex, pdftitle={Article}, pdfauthor={Author}]{hyperref} 
\usepackage{color}
\usepackage{bbold}
\usepackage{tabu}
\usepackage{mathtools}

\DeclarePairedDelimiter\bra{\langle}{|}
\DeclarePairedDelimiter\ket{|}{\rangle}

\begin{document}
\title{
Resonances, Recurrence Times and Steady States in Monitored Noisy Qubit Systems
}

\author{S.~Ma}
\email{Corresponding author: shuangerma2021@gmail.com}
\affiliation{Department of Physics, Institute of Nanotechnology and Advanced Materials, Bar-Ilan University, Ramat-Gan 52900, Israel}

\author{S.~Tornow}
\email{Corresponding author: sabine.tornow@unibw.de}
\affiliation{Research Institute CODE, University of the Bundeswehr Munich, 81739 Munich, Germany}

\author{E.~Barkai}
\affiliation{Department of Physics, Institute of Nanotechnology and Advanced Materials, Bar-Ilan University, Ramat-Gan 52900, Israel}

\begin{abstract}

We study non-equilibrium steady states and recurrence times in noisy, stroboscopically monitored qubit systems using complete measurements. In the noiseless limit, recurrence times are integer-quantized, with dips to lower integers when sampling approaches revival conditions associated with ergodicity breaking. Using an IBM quantum platform, we find that quantization is robust when sampling far from revivals, but breaks down dramatically near revivals: even weak noise produces large deviations and can invert the expected dips into pronounced peaks. To explain this behavior, we formulate a statistical-physics model of monitored noisy circuits in which monitoring drives an effective infinite-temperature steady state while thermal-like relaxation competes to favor a low-temperature limit. We show that the sampling time tunes a crossover between these regimes, near revivals stabilizing low-temperature behavior, and far from revivals restoring infinite-temperature behavior— with noise strength and detuning acting as coupled small parameters near resonance.

\end{abstract}

\maketitle

\section{Introduction}

Classical and quantum recurrence times arise across many disciplines, ranging from chaos theory and the foundations of statistical physics to, more recently, qubit systems and quantum circuits \cite{Kac,Kac1960,Poincare,BocchieriLoinger,Altman,Grunbaum2013,Grunbaum2019,Friedman,YinLarge,DidiBarkai,TornowZiegler,YinPNAS,WangEntropy,Stefanak2026recurrencein,Nitsche,Sinkovicz2015,Walter2025,Sinkovicz,Karimi}. In quantum mechanics, recurrence theory states that an isolated, energy-preserving system with a discrete spectrum returns arbitrarily close to its initial state after sufficiently long times \cite{BocchieriLoinger}. This result extends the classical Poincaré recurrence theorem \cite{Poincare}, which guarantees recurrence in finite classical phase spaces. These recurrence theorems are closely related to revival phenomena: under unitary dynamics, there may exist times at which the return probability satisfies $|\langle \psi(t)\vert\psi(0)\rangle|^2=1$ \cite{AverbukhPerelman,Robinett}. In such settings, recurrence is deterministic and was historically invoked to challenge Boltzmann's ideas of ergodicity and relaxation to steady states \cite{Ehrenfest}. For macroscopic systems, however, recurrence times of isolated dynamics are typically so long as to be physically irrelevant. In contrast, for small qubit systems, such as those realized on present-day noisy intermediate-scale quantum devices, deterministic recurrence scales can become experimentally accessible, as we demonstrate below.

For classical stochastic dynamics \cite{Norris,vanKampen,Redner2001}, recurrence is naturally formulated in probabilistic terms. For discrete time ergodic Markov processes, Kac's lemma \cite{Kac,Kac1960,Aldous,Kemeny,Condamin} relates the mean recurrence time to a given state $i$ to its steady state probability,
\begin{equation}
\langle n(i) \rangle = \frac{1}{p_i^{\rm ss}},
\label{eq01}
\end{equation}
where $p_i^{\rm ss}$ is the steady state probability and $\langle n(i)\rangle$ denotes the mean number of time steps until recurrence. This relation connects a dynamical quantity, the recurrence time, to a static steady state property of the system, provided the dynamics is ergodic. For example, in a random walk on a finite graph, the steady state occupation probabilities alone determine the mean recurrence times to the nodes of the graph. Importantly, in classical systems measurements are typically non-invasive, so that both steady states and mean recurrence times are insensitive to the measurement process.

Recently, monitored quantum recurrence has emerged as an active area of research \cite{Grunbaum2013,Grunbaum2019,Friedman,YinLarge,DidiBarkai,TornowZiegler,YinPNAS,WangEntropy,Stefanak2026recurrencein,Nitsche,Sinkovicz2015,Sinkovicz,Walter2025,Stefanak2026recurrencein,rbtb-8d27,Karimi}. In this framework, a quantum system undergoes unitary evolution interrupted by stroboscopic projective measurements \cite{Dhar,VecchioDelVecchio2026,ladenburger2025,Finocchiaro2025}, which generate a detection record from which recurrence times are defined. Unlike in classical dynamics, measurements fundamentally alter the system's evolution \cite{Skinner,GoogleQuantumAI2023,Koh2023}, rendering recurrence statistics sensitive to the sampling interval and the measurement protocol \cite{WisemanMilburn}. When the measurements are complete \cite{vonNeumann}, as defined below, quantum coherences are removed at each measurement, while the recurrence time defined through repeated monitoring remains inherently stochastic \cite{Strasberg,Smirne}. Remarkably, in ideal noise-free settings, the interplay between unitary dynamics and monitoring gives rise to quantization of the mean recurrence time: the mean number of measurements $\langle n \rangle$ assumes integer values and undergoes abrupt transitions between them as the sampling time is varied \cite{DidiBarkai}. Within this ideal theory, sampling at revival times can also lead to pronounced dips to lower integers, reflecting measurement-induced ergodicity breaking tied to the underlying unitary dynamics \cite{DidiBarkai,soldner2025}.

The key issue underlying the steady states and recurrence properties of realistic, noisy quantum platforms is the coexistence of two distinct Hamiltonians. The first is generated by a synthetic (engineered) Hamiltonian $H_{\rm syn}$, implemented through a sequence of quantum gates and governing the intended unitary dynamics. In contrast, thermal relaxation originates from the physical device Hamiltonian $H_{\rm phys}$, which is fixed by the hardware qubits and their coupling to the environment \cite{Martinis2005,Koch2007,Krantz2019,Reimann,Lvov,Andersen2025,Tuokkola2025,Cech,Hothem2025}. The interplay of these two Hamiltonians, together with repeated measurements, generally drives the system into a non-equilibrium steady state \cite{Gotta2026}, defined as the stationary distribution of measurement outcomes, whose properties we aim to characterize.

As we show, in realistic noisy qubit platforms our observations cannot be explained by deterministic Poincaré-like revivals, neither by a measurement-free steady-state description, nor by the noise-free quantization picture taken in isolation, even though each remains insightful in its respective limit. Using an IBM quantum processor, we demonstrate that at sampling intervals where the underlying unitary dynamics revives, recurrence is not achieved within one step; instead, it becomes a genuinely stochastic event and can exhibit extreme sensitivity even to weak noise. As a consequence, dips in recurrence times predicted by ideal monitored quantum dynamics \cite{Grunbaum2013,Friedman,YinLarge,DidiBarkai,TornowZiegler,YinPNAS,WangEntropy,Grunbaum2019} can be strongly suppressed or even inverted into peaks in the mean recurrence time. Moreover, the recurrence statistics need not respect the time reversal symmetry of the ideal theory, consistent with a physical bias toward relaxation into the qubit ground state. Finally, because the act of monitoring imprints an explicit $\tau$ dependence on the effective dynamics, recurrence cannot be inferred from a $\tau$ independent steady state. These observations motivate a theory of monitored recurrence in noisy quantum devices, which we develop in this work to identify noise mechanisms consistent with the data and to reveal how the sampling time controls the non-equilibrium steady state and associated recurrence statistics.

\section{Results}

\subsection{Recurrence times on remote quantum processors} 

In this section, we use remote experiments \cite{Georgescu,Li2023} on IBM quantum processors. The role of these demonstrations is to expose regimes in which existing noise-free recurrence theories (to be explained) break down and to guide the construction of effective theoretical descriptions. In particular, we observe large deviations from the integer mean recurrence time predicted by ideal theory. These deviations are most pronounced near special sampling intervals related to revival times.

 \textit{Idealized noiseless model.} 
We consider a finite-dimensional Hilbert space spanned by qubit states $\ket{\psi}$ in a quantum processor. For a system of $N$ qubits, the Hilbert space dimension is $2^N$. In the absence of decoherence, the dynamics is generated by a time-reversal-symmetric Hamiltonian $H_{{\rm syn}}$, with unitary evolution $U(\tau)=\exp(-iH_{{\rm syn}}\tau)$ (setting $\hbar=1$), implemented in practice using quantum gates. To define recurrence, we perform complete projective measurements at fixed time intervals $\tau$, which destroy quantum coherence. At each measurement, all $N$ qubits are measured in the computational basis $\{\ket{0},\ket{1}\}$. A designated target state---chosen to coincide with the initial state---is monitored, and detection occurs at the \emph{first} measurement time $n\tau$ at which this state reappears. The integer $n$ is a random variable defining the recurrence time.

Since each measurement yields a classical bit string, for example $\ket{1100\cdots 10}$, both the initial and target states are product states in the $z$ basis. As a concrete illustration, consider a three-qubit system ($N=3$), for which the Hilbert space contains $8$ states. The system is prepared initially in the state $\ket{000}$, which also serves as the target state. After the first unitary evolution and measurement, the state may be recorded as $\ket{100}$ at time $\tau$. Following a second unitary step, the measurement at time $2\tau$ may yield $\ket{011}$. If the target state $\ket{000}$ is observed at time $3\tau$, the protocol terminates. In this realization, the recurrence time is $n=3$ in units of $\tau$.

Repeating this protocol over many realizations yields the mean recurrence time $\langle n \rangle$, which is an integer-valued quantity \cite{DidiBarkai}. In general, it is bounded by
\[
1 \le \langle n \rangle \le \dim(\mathcal{H_{{\rm syn}}}) .
\]
The lower bound $\langle n \rangle=1$ is attained when $\tau$ coincides with a revival time, namely when $U(\tau)$ is diagonal. In this case, the system returns to its initial state after a single evolution step and is detected with unit probability at the first measurement. This identifies $\tau$ as a natural control parameter for recurrence, although in the ideal noise-free setting the mean recurrence time can change only through transitions between integer values. More specifically, the integer-valued quantity $\langle n\rangle$ counts the number of states linked by nonzero Born transition probabilities (defined below) within a generally fragmented Hilbert space \cite{DidiBarkai}.

\begin{figure}\begin{center}
\epsfig{figure=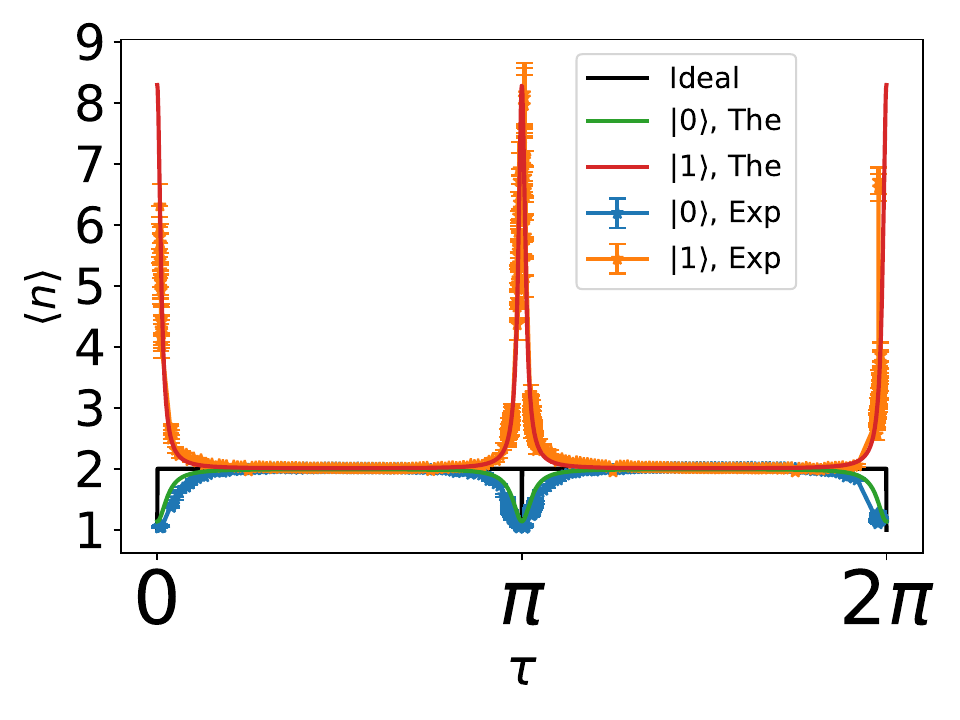
, width=0.45\textwidth,trim=0mm 0mm 0mm 0mm, clip}
\end{center}
\caption{
Mean recurrence time $\langle n \rangle$ for states $\ket{0}$ and $\ket{1}$ versus sampling time $\tau$ in a single IBM qubit system. 
Without noise, $\langle n \rangle = 2$ except at simple revival resonances, where $\langle n \rangle = 1$, as shown by the ideal black solid line. 
Near resonances, noise effects are pronounced, even if the noise per cycle is weak. 
Far from resonances the recurrence time for the $\ket{0}$ and $\ket{1}$ states
are identical and given by the integer $2$ as expected from noise-free theory
and time reversal symmetry.
At resonances, we observe time reversal symmetry breaking where quantization of the mean recurrence time is clearly invalid.   
The excited state $\ket{1}$ has a smaller population in a thermodynamics sense  and thus a longer recurrence time than the ground state $\ket{0}$. 
Theory from Eq.~(\ref{eq02}) matches the data quantitatively. A least-squares fit gives $p_{0\to1}=0.0012$ and $p_{1\to0}=0.0086$, with $R^2=0.941$. Error bars indicate the statistical uncertainty of the measured recurrence times. The experimental setup is described in the Methods section, and additional details of the data analysis are provided in Supplementary Materials S2.
}
\label{fig1}
\end{figure} 

\emph{A single qubit} provides the simplest platform for studying quantum recurrence and its deviation from the noiseless limit. We perform two independent remote experiments, one targeting return to $\ket{0}$ and the other to $\ket{1}$, both eigenstates of $\sigma_z$. The synthetic Hamiltonian is $H_{{\rm syn}} =\sigma_x$, so that the evolution between measurements is $U(\tau)=\exp(-i\sigma_x\tau)$. For an ideal noiseless qubit, it is straightforward to show that the mean recurrence time satisfies $\langle n\rangle=2$ for all $\tau$ except at $\tau=k\pi$ ($k$ integer), where $\langle n\rangle=1$ \cite{Grunbaum2013}. Thus, $\langle n\rangle$ is nearly constant at $2$, with sharp dips to $1$ at revival times when the initial state is detected with unit probability at the first measurement. Time-reversal symmetry under $\ket{0}\leftrightarrow\ket{1}$ ensures identical recurrence statistics in the ideal case, a hallmark of coherent and isolated evolution. 

Fig.~\ref{fig1} shows the corresponding results obtained on IBM quantum hardware. While $\langle n\rangle\approx2$ holds for most values of $\tau$, indicating the
anticipated integer mean recurrence time, pronounced deviations appear near the resonances and are accompanied by an explicit noise-induced breaking of the $\ket{0}\leftrightarrow\ket{1}$ symmetry: the $\ket{1}$ target exhibits a pronounced peak, whereas the $\ket{0}$ target shows a dip. At the peak, $\langle n\rangle$ exceeds the coherent bound set by the Hilbert space dimension ($2$). This behavior originates from transmon physics and coupling to the environment: in the absence of driving, $\ket{0}$ and $\ket{1}$ correspond to the ground and excited states, respectively, and energy relaxation favors $\ket{0}$. In a thermal limit, the higher steady-state 
probability of $\ket{0}$ implies a shorter recurrence time according to Kac's
lemma Eq.  
(\ref{eq01}).

The absence of detectable deviations far from resonances implies that the noise per evolution cycle is small. We model the behavior observed in Fig.~\ref{fig1} using noisy mid-circuit measurements, where each evolution cycle is accompanied by small incoherent relaxation probabilities $p_{1\to0}$ and $p_{0\to1}$. Since the thermal ground state of the physical qubit is $\ket{0}$, the relaxation probability $p_{1\to0}$ is typically larger than the reverse transition $p_{0\to1}$. The precise formulation of the noise model is presented in generality in the following sections. While the model considered here is deliberately simple, we have examined several alternative noise models using Kraus operators \cite{NielsenChuang2010}. Among them, the generalized amplitude-damping model provides consistent behavior, whereas bit-phase-flip, depolarization, and amplitude-damping models do not reproduce the observed features (see Supplementary Materials S4).

Physically, the central question is how such weak noise can nevertheless produce a very large effect near resonances. In particular, when $\tau$ is  close to revival times the noise can invert the sharp dips in the mean recurrence time predicted for the noiseless system into pronounced peaks when the target state is $\ket{1}$. Notably, $\ket{1}$ is not the excited state of the dynamical Hamiltonian $H_{{\rm syn}}=\sigma_x$. This highlights a key distinction between the thermal ground state of the physical qubit, $\ket{0}$, and the ground state of the effective Hamiltonian governing the unitary dynamics. This mismatch arises because the Hamiltonian $H_{{\rm syn}}$ is synthetically engineered through quantum gates rather than reflecting the intrinsic energy structure of the hardware.

In agreement with the data in Fig. \ref{fig1}, for recurrence to the state $\ket{0}$, the noisy measurement model, presented below, predicts
\begin{equation}
\langle n\rangle = 2 + \frac{2\left(p_{0\to 1} - p_{1\to 0}\right)}
{1 - p_{0\to 1} + p_{1\to 0} - \left(1 - p_{0\to 1} - p_{1\to 0}\right)\cos 2\tau},
\label{eq02}
\end{equation}
while for recurrence to $\ket{1}$ the result follows by exchanging
$p_{0\to 1} \leftrightarrow p_{1\to 0}$. When the noise is weak, away from resonance
($\cos 2\tau \neq 1$), $\langle n\rangle \approx 2$
consistent with the ideal coherent value. At resonance, however, one finds
\begin{equation}
\langle n\rangle_{\ket{0}} = \frac{p_{0\to 1} + p_{1\to 0}}{p_{1\to 0}}, \qquad
\langle n\rangle_{\ket{1}} = \frac{p_{0\to 1} + p_{1\to 0}}{p_{0\to 1}} .
\end{equation}
The asymmetry in the transitions is given by
the ratio  $p_{0\to 1}/p_{1\to 0} = \exp( - \Delta E/T_{{\rm  eff}})$ 
where $\Delta E$
represents an energy gap between the physical ground state and 
the excited state
and $T_{{\rm eff}}$ is an effective temperature \cite{BreuerPetruccione,GardinerZoller}. 
In our case $\exp(-\Delta E/T_{{\rm eff}}) = 0.14$,
indicating a strong bias toward the ground state, consistent with a low-temperature–like regime typical of superconducting quantum hardware.
When the sampling and the revival times are matching, 
the recurrence time  is strongly renormalized by weak noise,
and  the system becomes highly susceptible to incoherent relaxation. 


\begin{figure}\begin{center}
\epsfig{figure=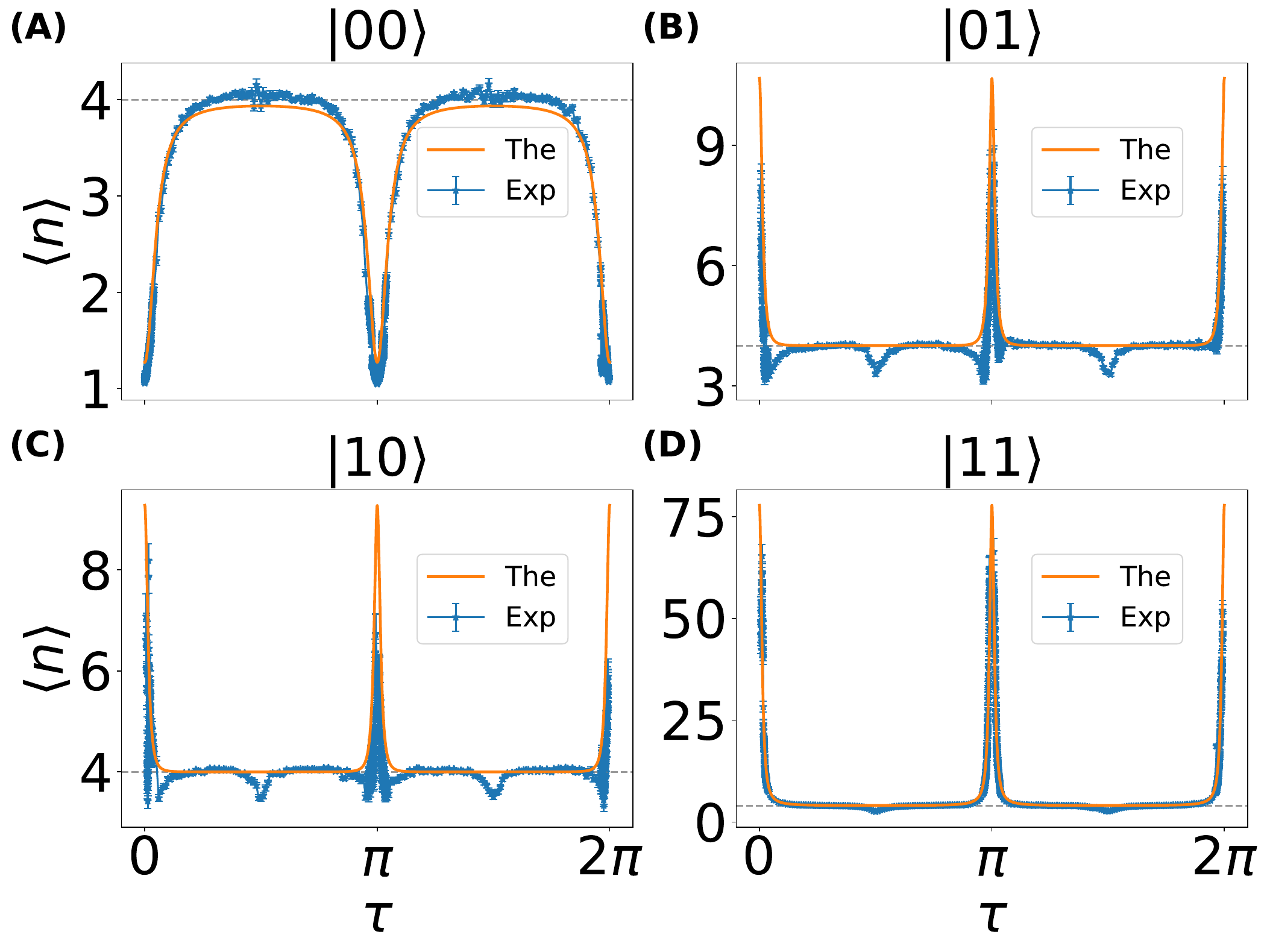, width=0.48\textwidth,trim=0mm 0mm 0mm 0mm, clip}
\end{center}
\caption{
Mean recurrence time $\langle n \rangle$ versus sampling time $\tau$ for a two-qubit system with complete measurements. 
In the noiseless case, $\langle n \rangle = 4$ except for two classes of resonances: $\tau = \pi/2 + k\pi$ and $\tau = k\pi$. 
Weak noise removes the first class, while the second ($\tau = k\pi$) survives with modified dips, peaks, and asymmetry between basis states (see subtitles), reflecting noise sensitivity near resonances and negligible effects far from them. 
At resonance noise effects are amplified, with $\ket{00}$ $(\ket{11})$ showing pronounced dips (peaks) due to its high (low) occupation probability, consistent with Kac's lemma. 
The dashed gray line indicates the reference value $4$.
A joint least-squares fit of the weak-noise theory to all four panels yields
$p_{1\to0,a}=0.0189$,
$p_{0\to1,a}=0.0026$,
$p_{0\to1,b}=0.0022$, and
$p_{1\to0,b}=0.0188$, with $R^2=0.939$. Detailed analysis of the experimental data is presented in the Supplementary Materials S2.
}
\label{fig2}
\end{figure}

{
\em Four-level system.} 
We turn to two qubits so now $N=2$  and study separately
 the recurrence to the four basis states $\ket{00}$, $\ket{01}$, $\ket{10}$, and $\ket{11}$.
 The Hamiltonian implemented on the quantum hardware is $H_{{\rm syn}}=\sigma_x \otimes I+I \otimes \sigma_x$. 
Recurrence is monitored independently for each state, and results are compared. 
  Without noise, the
recurrence to any state is statistically identical,
 and the  mean recurrence
is  $\langle n\rangle=4$, except at sharp
 resonance values of $\tau$ where $\langle n  \rangle = 2$
or $\langle n \rangle = 1$.   
As demonstrated in Fig. \ref{fig2}
 experiments confirm $\langle n\rangle=4$ far from resonance. 
Since all four states exhibit the same behavior far from resonances, 
 Kac's lemma is 
indicative of an infinite temperature-like
 limit where populations in all the states are identical (see more details below). 
From single-qubit results we expect strong noise sensitivity 
near resonance; here unlike the simple qubit case, two distinct classes emerge,
one that breaks symmetry and one that does not. 

First, consider  resonances at $\tau=k\pi$ ($k\ge0$),
namely sampling times corresponding to revival periods, where noiseless theory predicts a transition
from $\langle n \rangle =4$ to $\langle n \rangle = 1$.
 We see clearly that the various states are
non identical, 
 $\langle n\rangle$ shows an extremely strong peak for $\ket{11}$, a dip
for  $\ket{00}$ while the states $\ket{01}$ and $\ket{10}$ are intermediate.
As before, this occurs because the physical Hamiltonian makes $\ket{11}$ the excited state, and it is therefore the least populated. This asymmetry again reflects low-temperature-like physics, where the preference for occupying the physical ground state $\ket{00}$ is pronounced.

Second, the noiseless theory predicts additional resonances at $\tau=\pi/2+k\pi$ and a transition from $\langle n \rangle=4$ to $\langle n \rangle=2$. In the presence of noise, however, these resonances are washed out: the sharp dip feature disappears. Yet, unlike the resonances at $\tau=k\pi$, we do not 
observe significant variations between the recurrence 
times of the four initial states. Thus, while both classes of ideal
resonances are noise-sensitive, their fate under noise differs qualitatively. This points to a richer phenomenology than in the single-qubit case. Inspired by these results, we present in the next section a general theory of steady states and recurrences in the monitored process of $N$ qubit systems. 

\subsection{Steady states for monitored quantum dynamics}

With the remote experimental observations in mind, we construct a physical model for  $N$-qubit systems. The computational basis states are denoted
 by $\ket{b_1,b_2,\ldots,b_N}$, where $b_i\in\{0,1\}$ labels the eigenvalues of the $\sigma_z$ operator and at each cycle the system evolves unitarily under 
$U(\tau)=\exp(-iH_{{\rm syn}} \tau)$.
As mentioned, we perform projective measurements of all qubits in the $\sigma_z$ basis every $\tau$ units of time. The recorded outcome uniquely fixes the post measurement state, so that coherence are removed at each step.
The dynamics therefore admits a Markovian description \cite{DidiBarkai}.
Let $\ket{p}$ denote the probability vector immediately after the $n$-th measurement, with components $p(i)=\bra{i}p\rangle$ where $\ket{i}$ denoted a computational basis state.
In the absence of noise, transitions between basis states are governed by Born's rule,
\begin{equation}
G_{ji}=|\langle j|U(\tau)|i\rangle|^2,
\label{eq03}
\end{equation}
where $G$ is doubly stochastic and time-reversal symmetric, $G_{ij}=G_{ji}$. 
For most $\tau$ the corresponding steady state $G \ket{p} =\ket{p}$ 
 is uniform implying that the measurements drive the system to a high-temperature limit. Yet, it is precisely the breakdown of this behavior that is of interest, as already shown in Figs.~\ref{fig1} and \ref{fig2}.

Physical noise induces additional incoherent transitions, which we describe by a stochastic matrix \(G'\) with elements \(G'_{ji}=p_{i\to j}\) for \(i\neq j\) and \(G'_{ii}=1-\sum_{j\neq i}p_{i\to j}\), where \(p_{i \to j}\) is the probability for a noise-induced transition from state \(i\) to \(j\) during a single measurement cycle \cite{GardinerZoller}. We assume that noise acts independently on each qubit.
Thus each qubit is characterized by a transition probability per unit cycle
$p_{0\to 1,q}$ which is much smaller than $p_{1 \to 0,q}$. We extend the Boltzmann-statistics description from a single qubit to many qubits, $p_{0\to 1,q}/p_{1 \to 0,q} = \exp( - \Delta E_q / T_{{\rm eff}})$. Our quantum computer demonstrations show that
qubits are not identical, and the index $q$ is a label for the qubit $q$.
The dynamics generated by \(G'\) alone relaxes the system
to a  thermal steady state \(\ket{p}\) which is determined by \(G'\ket{p}=\ket{p}\). This is the Boltzmann distribution associated with a physical Hamiltonian \(H_{\rm phys}\), where \(\langle i|p\rangle \propto e^{-E_i/(k_B T_{\rm eff})}\), where \(E_i\) is the energy of the computational basis state \(i\). Our remote experiments on cloud-based quantum processors are consistent, as a first approximation, with \(H_{\rm phys}=\sum_{q=1}^{N}\Delta E_q\,\sigma_z^{(q)}\), so that the physical ground state corresponds to all qubits being in the \(\ket{0}\) (spin-down) state.

The combined evolution over one measurement cycle is modeled with
a Markov matrix  $M=G'G$. We focus on the steady state
 $\lim_{n\to\infty}M^n\ket{p^{\rm initial}}=\ket{p^{\rm ss}}$, from which mean recurrence times follow via Kac's lemma Eq. (\ref{eq01}).
Since the measurements are strong and projective, the steady state depends explicitly on the sampling time $\tau$, so the recurrence times and steady-state properties acquires an intrinsic measurement dependence.
Importantly, if $G={\cal I}$ (the identity), the steady state is determined by $G'$ alone.
As mentioned for thermal relaxation,  this implies on qubit platforms a low-temperature-like limit, with a preferred population in the state $\ket{00...00}$.

We consider weak noise per cycle, $p_{ij}\ll1$, and write $G'=\mathcal{I}+\epsilon C$ with $\epsilon\ll1$. The steady state is expanded as $\ket{p^{\rm ss}}=\ket{p}_0+\epsilon\ket{p}_1+\cdots$. In the noiseless case 
\begin{equation}
\ket{p}_0=(1,1,\ldots,1)/2^N,
\label{eq04}
\end{equation}
corresponding to an infinite-temperature distribution induced by repeated measurements \cite{DidiBarkai,Yi}. An exception occurs at special sampling times when $G$ is 
block diagonal, indicating ergodicity breaking. 
The case $G=\mathcal{I}$ defines a revival time $\tau_R$, at which the unitary evolution reproduces the initial state and the noiseless dynamics is clearly non-ergodic even in the presence of the measurements. 
Away from revival, first-order perturbation theory yields
\begin{equation}
\ket{p}_1=\left(\mathcal{I}-G+\frac{J}{2^N}\right)^{-1}C\ket{p}_0,
\label{eq05}
\end{equation}
where $J$ is the all-ones matrix. This approach fails at revival times, where $G=\mathcal{I}$ and the inverse does not exist. At other singular points, such as $G^2=\mathcal{I}$, the approach remains valid arbitrarily close to the singularity, leading to a smooth continuation across these points. In contrast, higher-order relations $G^k=\mathcal{I}$ are incompatible with physically allowed time-reversal–symmetric dynamics, and thus never occur. The detailed derivation is shown in Supplementary Materials S5.

A distinct regime is therefore found close to revival times, where the perturbative term diverges. Writing $\tau=\tau_R+\delta\tau$, we identify two competing small parameters: the noise strength $\epsilon$ and the deviation $\delta\tau$. We seek a perturbative limit in which $\epsilon$ and $\delta\tau^2$ are of the same order. Expanding around revival yields $G(\tau)\simeq\mathcal{I}-(\delta\tau)^2W$, where 
\begin{equation}
W_{ij} =
\begin{cases}
- |H_{ij}|^2, & i \neq j \\
\sum_{k \neq i} |H_{ik}|^2, & i=j
\end{cases},
\end{equation}
a symmetric matrix determined by the Hamiltonian. Here, $H_{ij}$ denotes the matrix element of $H$ in the computational basis, i.e., $H_{ij}=\langle i|H|j\rangle$. To leading order, the steady-state condition reduces to
\begin{equation}
\bigl[\epsilon C-(\delta\tau)^2W\bigr]\ket{p^{\rm ss}}=0,
\label{eq06}
\end{equation}
so that $\ket{p^{\rm ss}}$ is the normalized null vector of $\epsilon C-(\delta\tau)^2W$. The steady state exactly at revival ($\delta\tau=0$) is noise dominated, implying
that close to revival sampling, the system will relax to a low-temperature-like limit. This explains the gigantic peaks observed in our demonstration on quantum platforms performing in a weak-noise limit.
The perturbative treatment near revival is substantially richer compared to the far from resonance case Eq. (\ref{eq05}): for finite $\delta\tau$ it captures a continuous crossover from a noise-controlled, low-temperature steady state to the infinite-temperature state obtained far from revival, where unitary-induced mixing restores the measurement-dominated dynamics. As a result, time-reversal symmetry is most strongly broken near revival times, where relaxation toward a low-temperature state dominates, while away from revival the dynamics are driven by the measurements toward an infinite-temperature steady state.

{\bf The Hypercube model.}
We use the $N$-dimensional hypercube as an analytically tractable $N$-qubit model with broad applications in quantum walks, quantum information processing, and multi-qubit dynamics \cite{ChildsGoldstone,MooreRussell,Kempe,KroviBrun,Howard2019,Goto2024}. Importantly for our purposes, its high symmetry exposes the sampling-time resonances of the unitary dynamics in a particularly transparent way and allows closed-form steady-state predictions in the presence of weak noise. Each vertex corresponds to a binary string $\ket{i}\equiv\ket{b_1 b_2 \ldots b_N}$ with $b_k=0,1$, and the Hamiltonian is $H=\sum_{k=1}^{N}\sigma_x^{(k)}$, where $\sigma_x^{(k)}$ denotes the Pauli-$X$ operator acting on qubit $k$. The transition matrix follows analytically,
\begin{equation}
G_{ji}(\tau)=\left(\cos^{2}\tau\right)^{N-d(i,j)}\left(\sin^{2}\tau\right)^{d(i,j)},
\end{equation}
where $d(i,j)$ is the Hamming distance between $\ket{i}$ and $\ket{j}$.

To incorporate noise, we assume independent and identical single-qubit noise acting on each qubit. This simplifying assumption is made to expose the underlying physical mechanisms governing the steady state. More realistic, device-specific noise can be treated numerically within the general framework developed above. The total noise matrix factorizes as $G' = (g')^{\otimes N}$ with
$g'=\bigl(\begin{smallmatrix} 1-p_{0\to1} & p_{1\to0}\\ p_{0\to1} & 1-p_{1\to0}\end{smallmatrix}\bigr)$,
$p_{0\to1}=\epsilon\gamma_{0\to1}$, and $p_{1\to0}=\epsilon\gamma_{1\to0}$.
In the weak-noise limit $\epsilon\ll 1$, $C=\sum_{k=1}^{N} c^{(k)}$ with
$c=\bigl(\begin{smallmatrix} -\gamma_{0\to1} & \gamma_{1\to0}\\ \gamma_{0\to1} & -\gamma_{1\to0}\end{smallmatrix}\bigr)$. Thus, the term $C\ket{p}_0$ appearing in Eq.~(\ref{eq05}) can be written as
\begin{equation}
C\ket{p}_0
=
\frac{\gamma_{1\to0}-\gamma_{0\to1}}{2^N}
\sum_{s\in\{0,1\}^N}
\bigl[N-2\boldsymbol{w}(s)\bigr]\ket{s},
\label{eqcp0}
\end{equation} where $\boldsymbol{w}(s)$ is the Hamming weight of $\ket{s}$. Due to the assumption of physically identical qubits, the steady-state result depends only on the measurement interval $\tau$, the noise level, and the Hamming weight of the recurrent state.

\begin{figure*}\begin{center}
\epsfig{figure=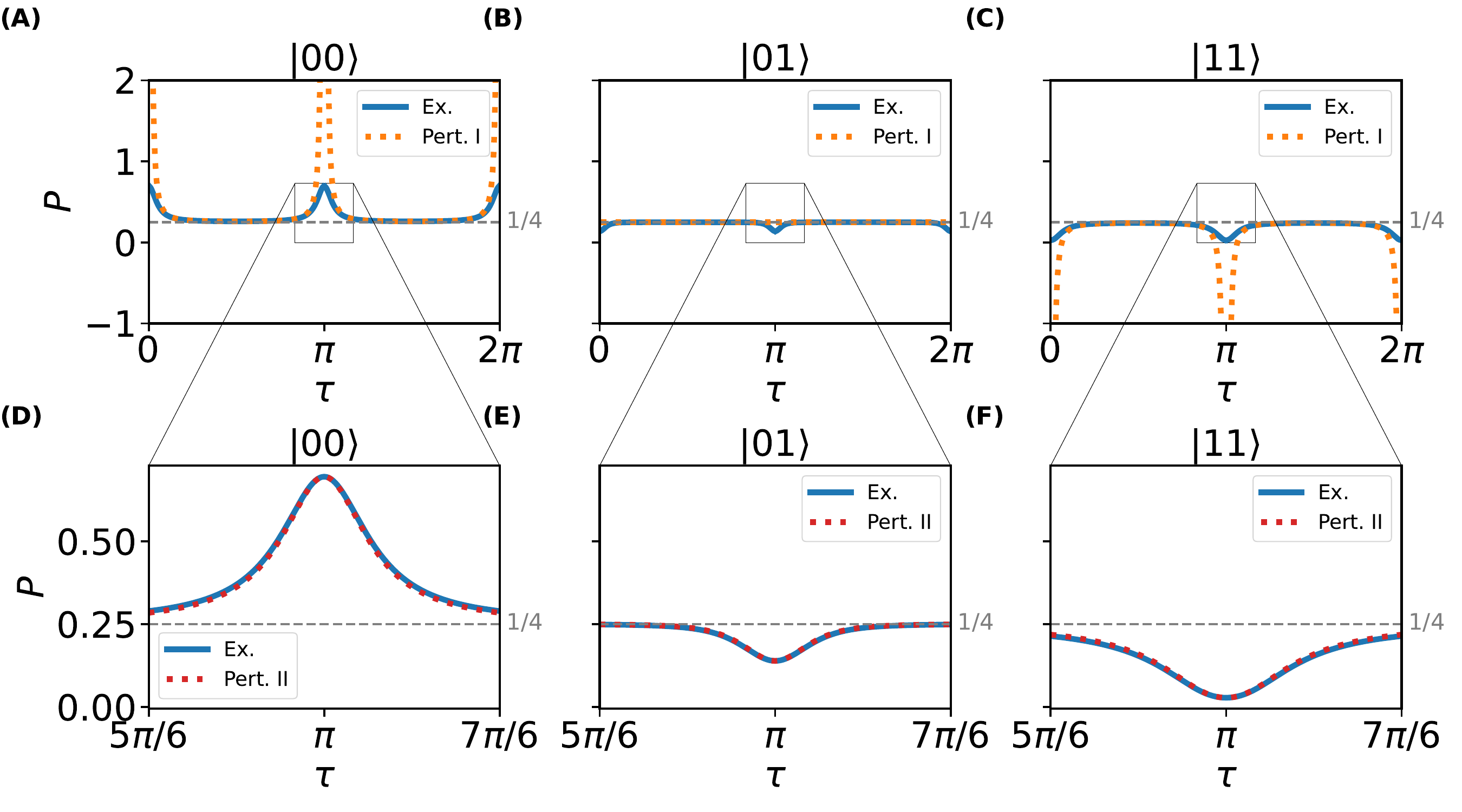, width=0.9\textwidth,trim=0mm 0mm 0mm 0mm, clip}
\end{center}
\caption{Steady-state probabilities grouped by the Hamming weight of the computational basis states of the Hypercube model. (A–C) Results from the first perturbation theory [Eq.~(\ref{eq05}, \ref{hyperI})], shown as orange dotted lines, are compared with exact numerical solutions (blue solid lines) for noise-induced transition probabilities $p_{1\to0}=0.05$ and $p_{0\to1}=0.01$. Excellent agreement is observed for generic sampling times $\tau$, except in the vicinity of $\tau=0,\pi,2\pi$, where noise-induced effects become non-perturbative. These special sampling times correspond to near-revival conditions with $G(\tau)\approx\mathcal{I}$, where the first perturbative expansion [Eq. (\ref{eq05})] breaks down. (D–F) Zoomed-in view near $\tau=\pi$. The red dotted lines show the predictions of the second perturbation theory [Eq. (\ref{eq06}, \ref{hyperII})], which treats both the noise strength and the deviation from $\tau_R$ as small parameters. In this resonant regime, the second perturbative approach [Eq. (\ref{eq06})] accurately captures the steady state and restores agreement with the exact numerical results (blue solid lines).}
\label{fig3}
\end{figure*}

Using Eq.~(\ref{eq04}, \ref{eq05}) together with the symmetry of the hypercube, the first-order correction to the steady state becomes
\begin{equation}
\ket{p}_1=\frac{1}{1-\cos(2\tau)}\,C\ket{p}_0.
\label{hyperI}
\end{equation}
Eq.~(\ref{hyperI}) immediately reveals the central message: away from revivals the correction is regular and accurately captures the steady state, while near revival times it breaks down in a controlled manner. In particular, for generic sampling times $\tau$ (including $\tau=\pi/2+k\pi$) the prefactor in Eq.~(\ref{hyperI}) stays finite, and the perturbative prediction matches the exact steady state. This success is demonstrated explicitly in Fig.~\ref{fig3}(A--C) for the two-qubit system, where the first perturbation theory agrees quantitatively with exact numerics.

In stark contrast, as $\tau$ approaches the revival times $\tau=k\pi$, the prefactor in Eq.~(\ref{hyperI}) diverges, and the perturbative expansion necessarily fails. Physically, these are resonant points of the underlying unitary dynamics, where the impact of weak noise is amplified by coherent revivals. This regime motivates a second perturbative description Eq. (\ref{eq06}) that is constructed specifically to remain accurate in the vicinity of revivals. 

Assuming independent and identical single-qubit noise acting on each qubit, the steady state close to revival times follows from Eq.~(\ref{eq06}) and reads
\begin{equation}
\langle i|p^{\mathrm{ss}}\rangle
=
\frac{
\left(p_{0\to1}+\delta\tau^{2}\right)^{\boldsymbol{w}}
\left(p_{1\to0}+\delta\tau^{2}\right)^{N-\boldsymbol{w}}
}{
\left(p_{0\to1}+p_{1\to0}+2\delta\tau^{2}\right)^{N}
},
\label{hyperII}
\end{equation}
where $\boldsymbol{w}$ is the Hamming weight of $\ket{i}$. As shown in Fig.~\ref{fig3}(D--F), Eq. (\ref{hyperII}) captures the steady state precisely where Eq. (\ref{hyperI}) fails, thereby providing a unified description across generic sampling times and resonant revival points. The hypercube thus yields closed-form $N$-qubit predictions, extending the analytical scope and applicability of our theory. Further examples and detailed analysis of a 4-dimensional hypercube are provided in the Supplementary Materials S3. Importantly, Fig.~\ref{fig3} also exposes the underlying physics of the resonant regime: weak noise near revival breaks the symmetry of the noiseless dynamics and produces strong bias toward low-Hamming-weight states, showing a low-temperature--like tendency.

{\bf Theory explains observations}. 
Our theoretical framework provides a consistent interpretation of the observations obtained from remote experiments on IBM quantum processors. Away from resonances, the dynamics is governed to leading order by the high-temperature phase described by Eq.~(\ref{eq04}). In this regime the steady state is uniform, and Kac's lemma predicts a mean recurrence time $\langle n\rangle = 2^{N}$ for all basis states, in agreement with what is observed. At resonance, when $G=\mathcal{I}$, the system instead enters a low-temperature–like regime in which recurrence times become strongly state dependent. In particular, recurrences to the states $\ket{0}$ (single qubit) and $\ket{00}$ (two qubits) are much faster than those to other states, as shown in Fig.~(\ref{fig1}) and Fig.~(\ref{fig2}). More specifically, for the single-qubit case we have $G=\bigl(\begin{smallmatrix}
\cos^2(\tau) & \sin^2(\tau) \\
\sin^2(\tau) & \cos^2(\tau)
\end{smallmatrix}\bigr)$ and $G'=\bigl(\begin{smallmatrix}
1 - p_{0\to 1} & p_{1\to 0} \\
p_{0\to 1} & 1 - p_{1\to 0}
\end{smallmatrix}\bigr)$, while for the two-qubit system $G=\bigl(\begin{smallmatrix}
\cos^2(\tau) & \sin^2(\tau) \\
\sin^2(\tau) & \cos^2(\tau)
\end{smallmatrix}\bigr)\otimes \bigl(\begin{smallmatrix}
\cos^2(\tau) & \sin^2(\tau) \\
\sin^2(\tau) & \cos^2(\tau)
\end{smallmatrix}\bigr)$ and $G'=\bigl(\begin{smallmatrix}
1 - p_{0\to 1,a} & p_{1\to 0,a} \\
p_{0\to 1,a} & 1 - p_{1\to 0,a}
\end{smallmatrix}\bigr)\otimes\bigl(\begin{smallmatrix}
1 - p_{0\to 1,b} & p_{1\to 0,b} \\
p_{0\to 1,b} & 1 - p_{1\to 0,b}
\end{smallmatrix}\bigr)$. The corresponding noise parameters are specified in the figure captions. For these parameters, the weak-noise perturbation theory, Eq.~(\ref{hyperI}) and Eq.~(\ref{hyperII}), captures the main features of the measured $\langle n\rangle$, including both deep dips and large peaks.

We also discussed the noiseless resonance of the two-qubit system at $\tau=\pi/2+k\pi$, where the idealized theory predicts a transition between integer-valued recurrence times $\langle n\rangle=4$ and $\langle n\rangle=2$. This transition is not observed due to noise. It corresponds to the case $G^{2}=\mathcal{I}$, which predicts relaxation toward the uniform steady state, in contrast to the $G=\mathcal{I}$ resonance. As a consequence, the observed recurrence time remains $\langle n\rangle\simeq 4$. Nonetheless, this sampling time is of conceptual interest: in the noiseless limit, the condition $G^{2}=\mathcal{I}$ implies a decomposition of the dynamics into two disconnected sectors, signaling ergodicity breaking. Noise restores ergodicity, but the approach to the ergodic steady state is slow, reflecting the weak noise per cycle characteristic of present-day quantum hardware.

\section{Discussion}

Mid-circuit measurements on quantum processors provide a concrete experimental realization of monitored quantum recurrence in finite-dimensional systems. Such measurements are widely used for state readout, error characterization, and control. In the specific context of recurrence, the integer-valued mean recurrence time plays a special role: deviations from integer values constitute a sensitive diagnostic of non-unitary effects, such as coupling to an environment, or departures from ideal projective measurement postulates.

Earlier experimental studies of quantum recurrence were constrained by hardware limitations and typically involved only a few tens of measurements on two- and three-qubit systems. In this regime, noise effects were largely negligible. Here, we take advantage of recent advances in quantum-computing platforms, in particular the availability of mid-circuit measurements on IBM devices enabling up to thousands of measurement cycles (see details in Methods). This capability allows us to access regimes dominated by measurement back-action and noise that were previously inaccessible. A key technical ingredient of our approach is the threading method, which makes it possible in practice to explore an effectively unlimited number of measurements; details are provided in the Methods Section.

We studied the effect of repeated measurements on recurrence times and steady states in noisy quantum circuits. Remote experiments were used to construct realistic noise models based on device calibration data. Although several models were considered (see Supplementary Materials S4), most failed to reproduce the observed behavior. For example, depolarization suppresses the resonances entirely without the appearance of peaks or dips, in contrast to observations. Similarly, a randomization of the time between measurements leads to analogous qualitative effects \cite{YinPNAS,Kessler2021}. In contrast, a strongly asymmetric relaxation toward the physical ground state captures the observations qualitatively well and therefore serves as the basis of our theoretical description.

Repeated projective measurements collapse the state onto computational basis states, destroying coherence and leading naturally to an effective post-measurement Markovian description. Despite its simplicity, this framework yields rich behavior. In particular, the steady state and hence the mean recurrence time—depends sensitively on the sampling time $\tau$, in contrast to classical systems subject to weak, non-invasive measurements. Far from resonances, the system approaches an effective high-temperature limit in which all basis states are approximately equally populated, yielding $\langle n \rangle \simeq 2^N$, with small perturbative corrections that can be computed analytically, see Eq. (\ref{eq05}).

In the absence of noise, unitary dynamics combined with repeated measurements can lead to ergodicity breaking at special sampling times, where the effective transition matrix becomes block diagonal. The dynamics is then confined to disconnected sectors of Hilbert space. Since each sector is smaller than the full space, recurrence times within a sector are shorter than in the ergodic case, producing the dips in the mean recurrence time predicted by the noise-free theory. Weak noise qualitatively modifies this picture. Although noise-induced transitions between sectors may be rare, once such a transition occurs, returning to the original sector typically requires another rare noise event. These long excursions strongly affect recurrence statistics and render both recurrence times and steady states extremely sensitive to weak noise.

The noise sensitivity is most pronounced when the sampling time is close to the unitary revival time. While the location of the resonance is fixed by coherent quantum dynamics, the corresponding value of the mean recurrence time is controlled primarily by noise. In this regime, two small parameters govern the dynamics: the deviation of $\tau$ from the revival time and the strength of the weak noise. This motivates a second perturbative description Eq. (\ref{eq06}), which accurately captures the behavior near resonance and connects smoothly to the regimes far from resonance.

Close to resonance, asymmetric relaxation leads to a steady state dominated by the physical ground state. Recurrence to this state is fast, while returns to excited states are typically much longer. This regime is naturally described by low-temperature–like physics associated with decay toward the ground state of the qubits. By varying the sampling time $\tau$, one can induce a crossover from a high-temperature-like regime for $\tau$ below the revival time, to low-temperature behavior near resonance, and back to a high-temperature-like regime for larger $\tau$. In this sense, the stroboscopic sampling time acts as a control parameter for the effective temperature of the monitored quantum system.

Taken together, these observations show that, while quantized recurrence and resonance dips are well understood in ideal and unital settings, realistic non-unital monitored dynamics display qualitatively different behavior, and their characterization under mid-circuit monitoring has remained a significant challenge in quantum hardware. By combining theory and experiment, our work addresses this challenge and, for complete measurements, establishes an effective model that not only explains the observed resonance-enhanced regime but also provides a framework for realistic monitored quantum dynamics beyond the present work.

\section{Methods}
\subsection{Monitored two-level dynamics on a single qubit}
\label{subsec:single_qubit}

We implement the monitored dynamics of a two-level system using a single qubit with computational basis states $\{\ket{0},\ket{1}\}$. Coherent evolution is generated by the tight-binding Hamiltonian $H_\text{syn}=\sigma_x$, so that a stroboscopic step of duration $\tau$ is
$
U(\tau)=\exp(-iH_\text{syn}\tau)=\exp(-i\sigma_x\tau).
$
On the processor this evolution is realized by a single-qubit rotation about the $x$ axis,
$
R_x(\theta)=\exp\!\left(-i\frac{\theta}{2}\sigma_x\right),
\label{rx}
$
with the identification $\theta=2\tau$. After each application of $R_x(\theta)$, we perform a complete projective measurement in the computational basis, thereby generating a discrete-time monitored trajectory. To probe recurrence to a target state (here $\ket{1}$), we initialize the qubit in $\ket{1}$ and repeat the sequence of unitary evolution and measurement at fixed intervals $\tau$. Each realization yields a binary record of outcomes, e.g.\ $0,0,0,1$, where the first detection of $\ket{1}$ occurs at the fourth attempt. We define the detection time $n$ as the number of measurement attempts required for the first successful detection and extract the mean recurrence time $\langle n\rangle$, expressed in units of $\tau$ \cite{WangEntropy, TornowZiegler}.

A practical limitation of current superconducting quantum processors is the finite number of mid-circuit measurements within a single circuit execution. In our experiments, this restricts each trajectory to a maximum of $1000$ stroboscopic measurement attempts per shot. The results are demonstrated in the Supplementary Materials S1. Although $1000$ measurements per shot suffice to show the resonance structure, this finite measurement budget remains a significant constraint near resonant sampling times, where convergence to the asymptotic result is markedly slow, so truncation removes late detection events and induces systematic bias in the inferred recurrence statistics \cite{YinPNAS}. To overcome this hardware-imposed horizon, we develop a threading protocol that effectively extends the measurement record by concatenating circuit segments while preserving the underlying stroboscopic dynamics.

\subsection{Threading method}
\label{subsec:threading}

To overcome the finite-depth limitation while remaining within experimentally accessible circuits, we introduce a post-processing protocol that we refer to as the \emph{threading method}. The key requirement is that each projective measurement collapses the system to a known basis state, so that subsequent evolution depends only on the last recorded outcome and not on unobserved coherences. For complete measurements, like those considered here, this condition is naturally satisfied.

\begin{figure}[!htbp]
    \centering
    \includegraphics[width=1\linewidth]{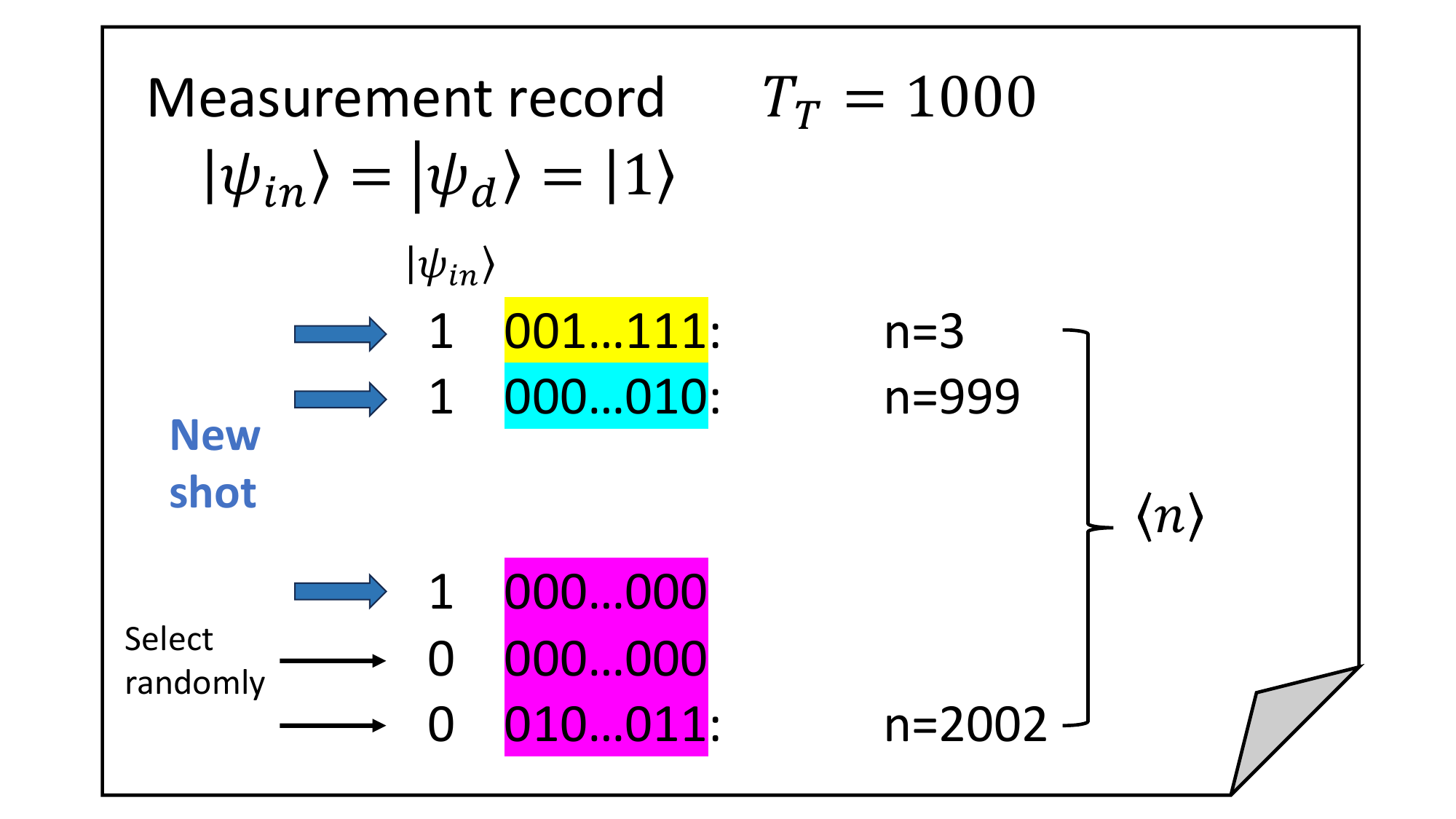}
    \caption{Examples of detection times for initial state $\ket{\psi_{in}}=\ket{1}$ with ``threading'' measurement protocol. Each trial consists of up to \( T_T = 1000 \) measurement attempts, a limitation that was imposed by the hardware. If the target state is detected, like in the first two records, the detection step is directly recorded. If detection fails, a new measurement sequence is initiated using the final state of the previous sequence and is randomly selected from the pool of trials. This process is iterated until we successfully detect the target state. In this way, the practical limitation of a finite number of measurements (1000 here) is bypassed, allowing us to observe new phenomena related to the recurrence time. The mean detection time \( \langle n \rangle \) is obtained by averaging over all successful events. In our experiments, we collected 40,000 measurement sequences, each of length 1000, on the IBM quantum processor ibm\_fez.
}
    \label{fig:methodschematic}
\end{figure}

For fixed parameters, we acquire a large ensemble of measurement sequences (threads), each consisting of $T_T$ stroboscopic steps. Starting from the chosen initial state, we scan a thread step by step. If the target state is detected within the first $T_T$ measurements, we record the corresponding step and terminate the procedure. If the target is not detected within $T_T$ measurements, we take the final measured state of that thread as the initial state for the next stage and select a new thread uniformly at random from the pool. This procedure is iterated until the target state is detected. In the experiment, we impose an upper limit of $10{,}000$ steps to limit the accumulation of measurement noise.

The first recurrence time $n$ is the cumulative number of measurement attempts over all concatenated threads up to the first detection event. The mean recurrence time $\langle n\rangle$ is obtained by averaging over many such constructed detection events. Fig.~\ref{fig:methodschematic} provides illustrative examples of the threading procedure. In our experiments, we collected $40{,}000$ shots of length $T_T=1000$ on the IBM quantum processor ibm\_fez. The details of the quantum processor are presented in the Supplementary Materials S6.

\section{Acknowledgement}
This work was supported by the Israel Science Foundation (Grant No. 2311/25). We acknowledge the use of IBM Quantum services for this work. The views expressed are those of the authors and do not reflect the official policy or position of IBM or the IBM Quantum team.


\begin{thebibliography}{99}
\bibitem{Poincare}
H.~Poincar\'e,
{\em Sur le probl\`eme des trois corps et les \'equations de la dynamique},
{\em Acta Math.} {\bf 13}, 1--270 (1890).

\bibitem{Kac}
M.~Kac,
{\em On the notion of recurrence in discrete stochastic processes},
{\em Bull.~Am.~Math.~Soc.} {\bf 53}, 1002--1010 (1947).

\bibitem{Kac1960}
M.~Kac, G.~E.~Uhlenbeck, A.~R.~Hibbs, B.~van~der~Pol, and J.~Gillis,
{\em Probability and related topics in physical sciences}
(1960).

\bibitem{BocchieriLoinger}
P.~Bocchieri and A.~Loinger,
{\em Quantum recurrence theorem},
{\em Phys.~Rev.} {\bf 107}, 337--338 (1957).

\bibitem{Altman}
E. G. Altmann, H. Kantz
{\em Recurrence time analysis, long-term correlations, and extreme events}
{\em Phys. Rev. E} {\bf  71} , 056106
215	(2005).

\bibitem{Grunbaum2013}
F.~A.~Gr\"unbaum, L.~Vel\'azquez, A.~H.~Werner and R.~F.~Werner,
{\em Recurrence for discrete time unitary evolutions},
{\em Commun.~Math.~Phys.} {\bf 320}, 543--569 (2013).

\bibitem{Grunbaum2019}
F.~A.~Gr{\"u}nbaum, C.~F.~Lardizabal and L.~Vel{\'a}zquez,
{\em Quantum Markov Chains: Recurrence, Schur Functions and Splitting Rules},
{\em Annales Henri Poincar\'e} {\bf 21}, 189--239 (2019).

\bibitem{Friedman}
H.~Friedman, D.~A.~Kessler and E.~Barkai,
{\em Quantum walks: The first detected passage time problem},
{\em Phys.~Rev.~E} {\bf 95}, 032141 (2017).

\bibitem{YinLarge}
R.~Yin, K.~Ziegler, F.~Thiel and E.~Barkai,
{\em Large fluctuations of the first detected quantum return time},
{\em Phys.~Rev.~Res.} {\bf 1}, 033086 (2019).

\bibitem{DidiBarkai}
A.~Didi and E.~Barkai,
{\em Measurement-induced quantum walks},
{\em Phys.~Rev.~E} {\bf 105}, 054108 (2022).

\bibitem{TornowZiegler}
S.~Tornow and K.~Ziegler,
{\em Measurement-induced quantum walks on an IBM quantum computer},
{\em Phys.~Rev.~Res.} {\bf 5}, 033089 (2023).

\bibitem{YinPNAS}
R.~Yin, Q.~Wang, S.~Tornow and E.~Barkai,
{\em Restart uncertainty relation for monitored quantum dynamics},
{\em Proc.~Natl.~Acad.~Sci.~U.S.A.} {\bf 122}, e2402912121 (2025).

\bibitem{WangEntropy}
Q.~Wang, S.~Ren, R.~Yin, K.~Ziegler, E.~Barkai and S.~Tornow,
{\em First hitting times on a quantum computer: Tracking vs. local monitoring,
topological effects, and dark states},
{\em Entropy} {\bf 26}, 10 (2024).

\bibitem{Nitsche}
T.~Nitsche \emph{et al.},
{\em Probing measurement-induced effects in quantum walks via recurrence},
{\em Sci.~Adv.} {\bf 4}, eaar6444 (2018).

\bibitem{Sinkovicz2015}
P.~Sinkovicz, Z.~Kurucz, T.~Kiss and J.~K.~Asb{\'{o}}th,
{\em Quantized recurrence time in unital iterated open quantum dynamics},
{\em Phys.~Rev.~A} {\bf 91}, 042108 (2015).

\bibitem{Sinkovicz}
P.~Sinkovicz, T.~Kiss and J.~K.~Asb{\'{o}}th,
{\em Generalized Kac lemma for recurrence time in iterated open quantum systems},
{\em Phys.~Rev.~A} {\bf 93}, 050101 (2016).

\bibitem{Walter2025}
B.~Walter, G.~Perfetto, and A.~Gambassi,
{\em Thermodynamic phases in first detected return times of quantum many-body systems},
{\em Phys.~Rev.~A} {\bf 111}, L040202 (2025).

\bibitem{Stefanak2026recurrencein}
M.~{\v{S}}tefa{\v{n}}{\'{a}}k, V.~Poto{\v{c}}ek, {\.{I}}.~Yal{\c{c}}{\i}nkaya, A.~G{\'{a}}bris and I.~Jex,
{\em Recurrence in discrete-time quantum stochastic walks},
{\em Quantum} {\bf 10}, 1982 (2026).

\bibitem{Karimi}
B.~Karimi, X.~Wu, A.~N.~Cleland, and J.~P.~Pekola,
{\em Blueprint for experiments exploring the quantum recurrence theorem on a coupled multiqubit system},
{\em Phys.~Rev.~Res.} {\bf 8}, L012062 (2026).

\bibitem{rbtb-8d27}
S.~Roy, S.~Gupta, and G.~Morigi,
{\em Causality, localization, and universality of monitored quantum walks with long-range hopping},
{\em Phys.~Rev.~E} {\bf 112}, 044146 (2025).

\bibitem{AverbukhPerelman}
I. Sh. Averbukh, N. F. Perelman
{\em Fractional revivals: Universality in the long-term evolution of quantum wave packets beyond the correspondence principle dynamics}
{\em Phys. Lett. A} {\bf 139}, 449--453 (1989).

\bibitem{Robinett}
R.~W.~Robinett,
{\em Quantum wave packet revivals},
{\em Phys.~Rep.} {\bf 392}, 1--119 (2004).

\bibitem{Ehrenfest}
P. Ehrenfest, T. Ehrenfest
{\em The Conceptual Foundations of the Statistical Approach in Mechanics}
(Dover Publications, 2014).

\bibitem{Norris}
J. R. Norris
{\em Markov Chains}
(Cambridge University Press, 1997).

\bibitem{vanKampen}
N. G. van Kampen
{\em Stochastic Processes}
in {\em Stochastic Processes in Physics and Chemistry}, 3rd ed.,
pp. 52--72 (Elsevier, 2007).

\bibitem{Redner2001}
S.~Redner,
{\em A Guide to First-Passage Processes}
(Cambridge University Press, 2001).

\bibitem{Aldous}
D.~Aldous and J.~A.~Fill,
{\em Reversible Markov Chains and Random Walks on Graphs}
(2002).

\bibitem{Kemeny}
J.~G.~Kemeny and J.~L.~Snell,
{\em Finite Markov Chains},
Van Nostrand, Princeton, NJ (1960).

\bibitem{Condamin}
S.~Condamin, O.~B\'enichou and M.~Moreau,
{\em First-passage times for random walks in bounded domains},
{\em Phys.~Rev.~Lett.} {\bf 95}, 260601 (2005).

\bibitem{Dhar}
S.~Dhar, S.~Dasgupta, A.~Dhar and D.~Sen,
{\em Detection of a quantum particle on a lattice under repeated projective measurements},
{\em Phys.~Rev.~A} {\bf 91}, 062115 (2015).

\bibitem{VecchioDelVecchio2026}
G.~Del~Vecchio~Del~Vecchio and S.~N.~Majumdar,
{\em Optimal detection of quantum states via projective measurements},
{\em J.~Phys.~A: Math.~Theor.} {\bf 59}, 035001 (2026).

\bibitem{ladenburger2025}
G.~Ladenburger, F.~Schmolke, and E.~Lutz,
{\em Universal first-passage time statistics for quantum diffusion},
arXiv:2511.03455 (2025).

\bibitem{Finocchiaro2025}
S.~Finocchiaro, G.~O.~Luilli, G.~Benenti, M.~G.~A.~Paris, and L.~Razzoli,
{\em Optimal quantum transport on a ring via locally monitored chiral quantum walks},
{\em Phys.~Rev.~E} {\bf 112}, 054116 (2025).

\bibitem{Skinner}
B.~Skinner, J.~Ruhman, and A.~Nahum,
{\em Measurement-Induced Phase Transitions in the Dynamics of Entanglement},
{\em Phys.~Rev.~X} {\bf 9}, 031009 (2019).

\bibitem{GoogleQuantumAI2023}
Google~Quantum~AI and Collaborators,
{\em Measurement-induced entanglement and teleportation on a noisy quantum processor},
{\em Nature} {\bf 622}, 481--486 (2023).

\bibitem{Koh2023}
J.~M.~Koh, S.~N.~Sun, M.~Motta \textit{et al.},
{\em Measurement-induced entanglement phase transition on a superconducting quantum processor with mid-circuit readout},
{\em Nat.~Phys.} {\bf 19}, 1314--1319 (2023).


\bibitem{WisemanMilburn}
H. M. Wiseman, G. J. Milburn
{\em Quantum Measurement and Control}
(Cambridge University Press, 2009).

\bibitem{vonNeumann}
J.~von~Neumann,
{\em Mathematical Foundations of Quantum Mechanics},
Princeton University Press (2018).

\bibitem{Strasberg}
P.~Strasberg and M.~G.~D\'iaz,
{\em Classical quantum stochastic processes},
{\em Phys.~Rev.~A} {\bf 100}, 022120 (2019).

\bibitem{Smirne}
A.~Smirne, D.~Egloff, M.~G.~D\'iaz, M.~B.~Plenio and S.~F.~Huelga,
{\em Coherence and non-classicality of quantum Markov processes},
{\em Quantum Sci.~Technol.} {\bf 4}, 01LT01 (2018).

\bibitem{soldner2025}
D.~Soldner, I.~Lesanovsky, and G.~Perfetto,
{\em Nonanaliticities and ergodicity breaking in noninteracting many-body dynamics via stochastic resetting and global measurements},
arXiv:2510.11450 (2025).

\bibitem{Martinis2005}
J.~M.~Martinis, K.~B.~Cooper, R.~McDermott, M.~Steffen, M.~Ansmann, K.~D.~Osborn, K.~Cicak, S.~Oh, D.~P.~Pappas, R.~W.~Simmonds, and C.~C.~Yu,
{\em Decoherence in Josephson Qubits from Dielectric Loss},
{\em Phys.~Rev.~Lett.} {\bf 95}, 210503 (2005).

\bibitem{Hothem2025}
D.~Hothem, J.~Hines, C.~Baldwin \textit{et al.},
{\em Measuring error rates of mid-circuit measurements},
{\em Nat.~Commun.} {\bf 16}, 5761 (2025).

\bibitem{Koch2007}
J.~Koch, T.~M.~Yu, J.~Gambetta, A.~A.~Houck, D.~I.~Schuster, J.~Majer, A.~Blais, M.~H.~Devoret, S.~M.~Girvin, and R.~J.~Schoelkopf,
{\em Charge-insensitive qubit design derived from the Cooper pair box},
{\em Phys.~Rev.~A} {\bf 76}, 042319 (2007).

\bibitem{Krantz2019}
P.~Krantz, M.~Kjaergaard, F.~Yan, T.~P.~Orlando, S.~Gustavsson, and W.~D.~Oliver,
{\em A quantum engineer’s guide to superconducting qubits},
{\em Appl.~Phys.~Rev.} {\bf 6}, 021318 (2019).

\bibitem{Reimann}
P.~Reimann,
{\em Typical fast thermalization processes in closed many-body systems},
{\em Nat.~Commun.} {\bf 7}, 10821 (2016).

\bibitem{Lvov}
D.~S.~Lvov, S.~A.~Lemziakov, E.~Ankerhold, J.~T.~Peltonen and J.~P.~Pekola,
{\em Thermometry based on a superconducting qubit},
{\em Phys.~Rev.~Appl.} {\bf 23}, 054079 (2025).

\bibitem{Andersen2025}
T.~I.~Andersen, N.~Astrakhantsev, A.~H.~Karamlou \textit{et al.},
{\em Thermalization and criticality on an analogue--digital quantum simulator},
{\em Nature} {\bf 638}, 79--85 (2025).

\bibitem{Tuokkola2025}
M.~Tuokkola, Y.~Sunada, H.~Kivij\"arvi, J.~Albanese, L.~Gr\"onberg, J.-P.~Kaikkonen, V.~Vestervinen, J.~Govenius, and M.~M\"ott\"onen,
{\em Methods to achieve near-millisecond energy relaxation and dephasing times for a superconducting transmon qubit},
{\em Nat.~Commun.} {\bf 16}, 5421 (2025).

\bibitem{Cech}
M.~Cech, I.~Lesanovsky, and F.~Carollo,
{\em Thermodynamics of Quantum Trajectories on a Quantum Computer},
{\em Phys.~Rev.~Lett.} {\bf 131}, 120401 (2023).

\bibitem{Gotta2026}
L.~Gotta, M.~Kulkarni, and G.~Perfetto,
{\em Towers of quantum many-body scars under stochastic resetting},
arXiv:2603.13165 (2026).

\bibitem{Georgescu}
I. M. Georgescu, S. Ashhab, F. Nori
{\em Quantum simulation}
{\em Rev. Mod. Phys.} {\bf 86}, 153--185 (2014).

\bibitem{Li2023}
H.~Li, Y.~Y.~Wang, Y.~H.~Shi \textit{et al.},
{\em Observation of critical phase transition in a generalized Aubry--Andr\'e--Harper model with superconducting circuits},
{\em npj~Quantum~Inf.} {\bf 9}, 40 (2023).

\bibitem{NielsenChuang2010}
M.~A.~Nielsen and I.~L.~Chuang,
{\em Quantum Computation and Quantum Information: 10th Anniversary Edition}
(Cambridge University Press, 2010).

\bibitem{BreuerPetruccione}
H.-P. Breuer, F. Petruccione
{\em The Theory of Open Quantum Systems}
(Oxford University Press, 2007).

\bibitem{GardinerZoller}
C. Gardiner, P. Zoller
{\em Quantum Noise: A Handbook of Markovian and Non-Markovian Quantum Stochastic Methods with Applications to Quantum Optics}
(Springer, 2004).

\bibitem{Yi}
J.~Yi, P.~Talkner and G.-L.~Ingold,
{\em Approaching infinite temperature upon repeated measurements of a quantum system},
{\em Phys.~Rev.~A} {\bf 84}, 032121 (2011).

\bibitem{ChildsGoldstone}
A.~M.~Childs and J.~Goldstone,
{\em Spatial search by quantum walk},
{\em Phys.~Rev.~A} {\bf 70}, 022314 (2004).

\bibitem{MooreRussell}
C.~Moore and A.~Russell,
{\em Quantum walks on the hypercube}
(2001).

\bibitem{Kempe}
J.~Kempe,
{\em Discrete quantum walks hit exponentially faster},
{\em Probab.~Theory Relat.~Fields} {\bf 133}, 215--235 (2005).

\bibitem{KroviBrun}
H.~Krovi and T.~A.~Brun,
{\em Hitting time for quantum walks on the hypercube},
{\em Phys.~Rev.~A} {\bf 73}, 032341 (2006).

\bibitem{Howard2019}
L.~A.~Howard, T.~J.~Weinhold, F.~Shahandeh, J.~Combes, M.~R.~Vanner, A.~G.~White, and M.~Ringbauer,
{\em Quantum Hypercube States},
{\em Phys.~Rev.~Lett.} {\bf 123}, 020402 (2019).

\bibitem{Goto2024}
H.~Goto,
{\em High-performance fault-tolerant quantum computing with many-hypercube codes},
{\em Sci.~Adv.} {\bf 10}, eadp6388 (2024).

\bibitem{Kessler2021}
D.~A.~Kessler, E.~Barkai, and K.~Ziegler,
{\em First-detection time of a quantum state under random probing},
{\em Phys.~Rev.~A} {\bf 103}, 022222 (2021).


\end{thebibliography}
\end{document}


\maketitle

\tableofcontents
\clearpage

\setcounter{figure}{0}
\setcounter{table}{0}
\setcounter{equation}{0}

\section{Experimental results with 1000 measurements}
\begin{figure}[!htbp]
    \centering
    \includegraphics[width=0.5\linewidth]{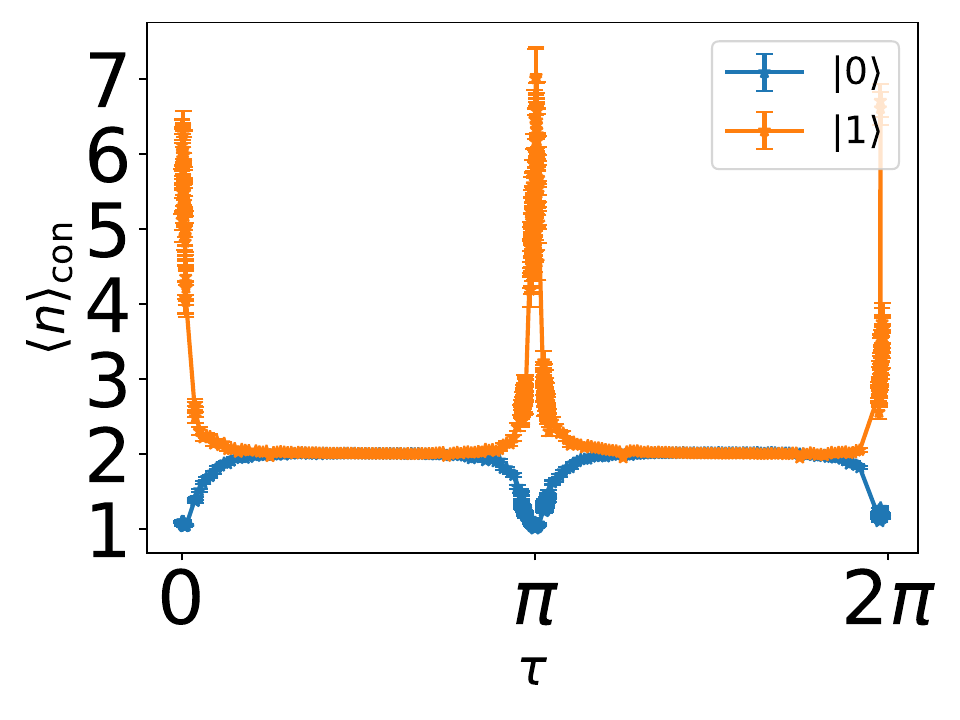}
    \caption{Conditional mean recurrence time $\langle n\rangle_{\rm con}$ for the single-qubit system with $T_R=1000$ stroboscopic measurement attempts. The curve follows the same overall trend as the threading-method analysis, but the resonance peak is reduced by the finite measurement limit. Error bars represent SEM of $\langle n\rangle_\text{con}$ at each sampling time~$\tau$.}
    \label{fig:1000real}
\end{figure}

In the main text, the experimental recurrence statistics are extracted using the threading protocol, which effectively extends the measurement record beyond the hardware limit (see Methods). In a single circuit run, each trajectory is restricted to at most $T_R=1000$ stroboscopic measurement attempts per shot on present IBM quantum computers. Occasionally, the target state is not detected up to the experimental cutoff of 1000 attempts. Such events are rare but become increasingly relevant in the vicinity of resonances \cite{YinPNAS}. Since the experiment records sequences only up to this maximal length, trajectories with no detection within the cutoff are excluded from the statistics. We define the mean recurrence time conditioned on detection
\begin{equation}
\langle n \rangle_{\mathrm{con}} =
\frac{\sum_{n=1}^{T_R} n\,F_n}{\sum_{n=1}^{T_R} F_n}\, .
\end{equation}
Here $F_n$ is the probability of first detection at the $n$-th measurement. Results obtained with 1000 repeated measurements are nevertheless sufficient to resolve the resonance peaks and dips. Figure~\ref{fig:1000real} displays the corresponding data obtained with $T_R=1000$, highlighting the resonance structure. 

\section{Data analysis and slow convergence}
In Figs.~\ref{fig1} and \ref{fig2}, we exclude data points in the immediate vicinity of the resonances. This is motivated by two effects: markedly slower convergence and enhanced fluctuations. We discuss both in detail below.

\subsection{Slow convergence near resonances}
\begin{figure}[!htbp]
    \centering
    \includegraphics[width=0.5\linewidth]{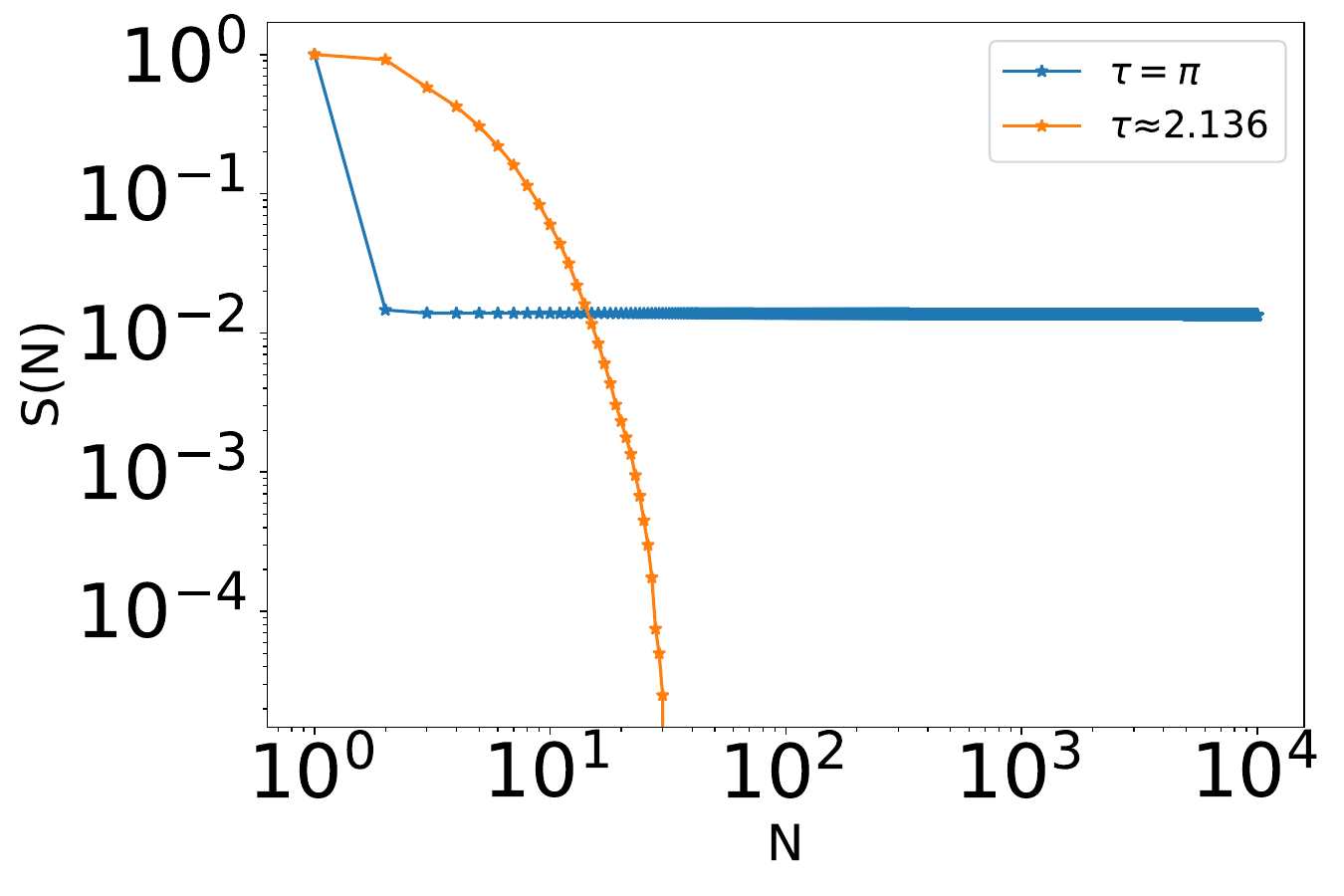}
    \caption{Survival probability of the recurrence time for the initial state $\ket{11}$, comparing a sampling point away from resonance ($\tau=2.136$) with the resonant case ($\tau=\pi$). The missing ratios are $0.0000$ for $\tau=2.136$ and $0.0134$ for $\tau=\pi$, respectively. The slow decay near resonance highlights slow convergence and pronounced long-tail effects.}
    \label{fig:CCDF}
\end{figure}

In the vicinity of resonance, convergence becomes markedly slow. Figure~\ref{fig:CCDF} shows the survival probability for the initial state $\ket{11}$, namely the probability that recurrence has not been detected up to step $N-1$. Explicitly, if $F_n$ denotes the probability that the first recurrence is detected at step $n$, then the survival probability is defined by
\begin{equation}
S(N) = 1 - \sum_{n=1}^{N-1} F_n.
\end{equation}
For $\tau=2.136$, which lies well away from resonance, the survival probability decays rapidly with $N$, indicating that most trajectories recur within a small number of steps; correspondingly, the missing fraction beyond $10{,}000$ measurements is negligible. In contrast, at $\tau=\pi$ the survival probability exhibits a much slower decay after the first step, signaling the presence of long recurrence times. As a consequence, a non-negligible fraction of $0.0134$ of trajectories remains undetected.

\subsection{Large fluctuations}
\begin{figure}[!htbp]
    \centering
    \includegraphics[width=0.5\linewidth]{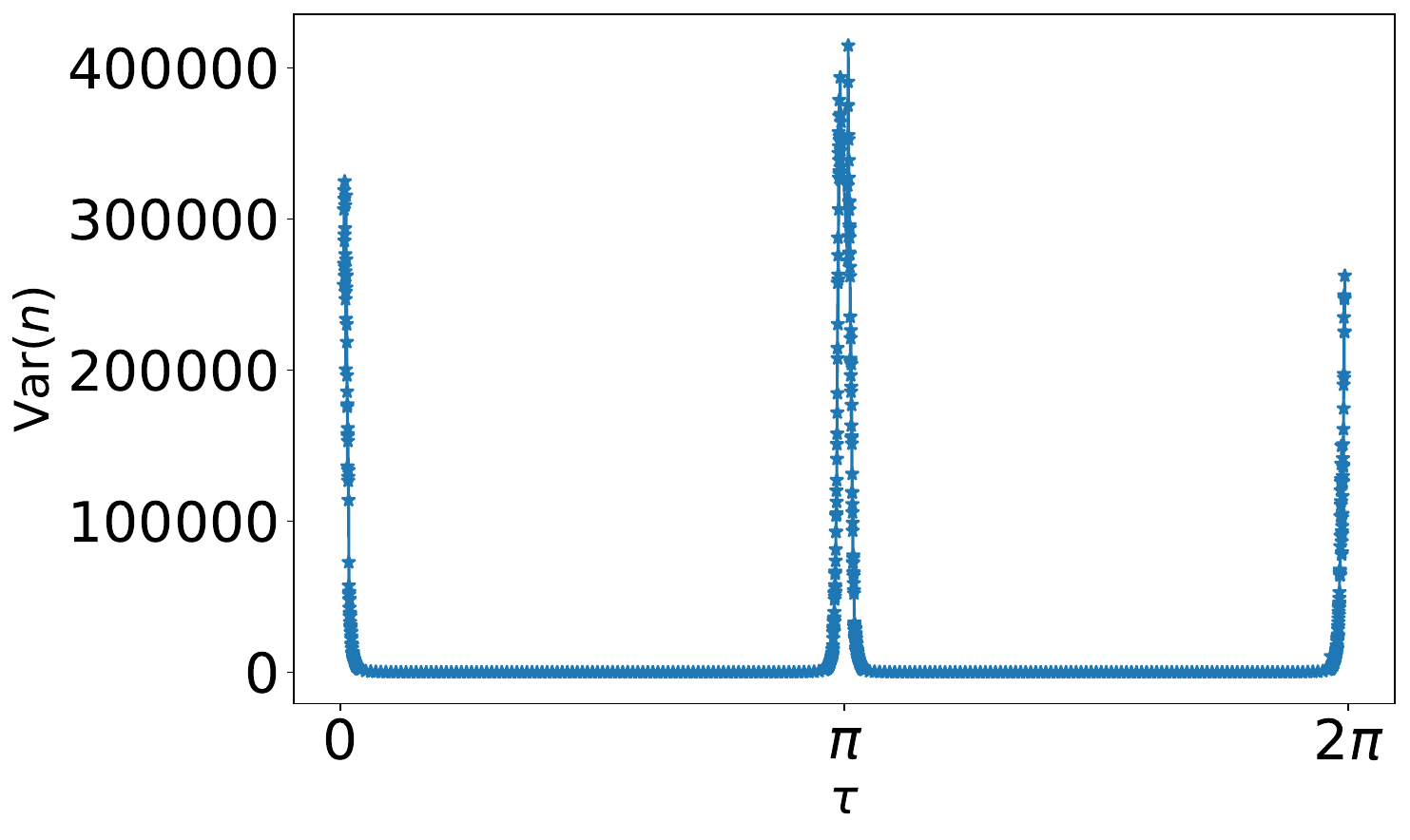}
    \caption{The sample variance of the recurrence time $n$ (in units of $\tau$) for the return to state $\ket{11}$, obtained using the experimental data with the threading method. Near resonance, the variance increases significantly, leading to noticeable fluctuations around the peak. Away from resonance, the variance remains small and the agreement between experimental and theoretical $\langle n \rangle$ is correspondingly tighter.}
    \label{fig:var}
\end{figure}
In the experiments, large fluctuations in $n$ are observed for the excited states, consistent with theoretical predictions \cite{Didi}. Fig.~\ref{fig:var} shows the experimental variance of $n$ for the system initialized in the state $\ket{11}$. Near resonance, the variance is strongly enhanced, leading to pronounced shot-to-shot fluctuations in the inferred peak height.
As a result, near resonance the finite-sample estimate of $\langle n \rangle$ becomes unreliable. 
In Figs.~\ref{fig1} and \ref{fig2}, we exclude data points in a narrow interval around the resonances at $\tau_R=0,\pi,2\pi$, namely those satisfying $|\tau-\tau_R|<\epsilon$ with $\epsilon=2\times10^{-2}$, where fluctuations are strongly enhanced and convergence becomes markedly slow.  Since this window is very small, the conclusions of the paper are unchanged. The slow convergence and large fluctuations are related to ergodicity breaking, which is found for the idealized noiseless case at resonances \cite{Didi}, as discussed in the text.

\section{Steady states of 4-dimensional hypercubes}
\begin{figure}[!htbp]
    \centering
    \includegraphics[width=0.5\linewidth]{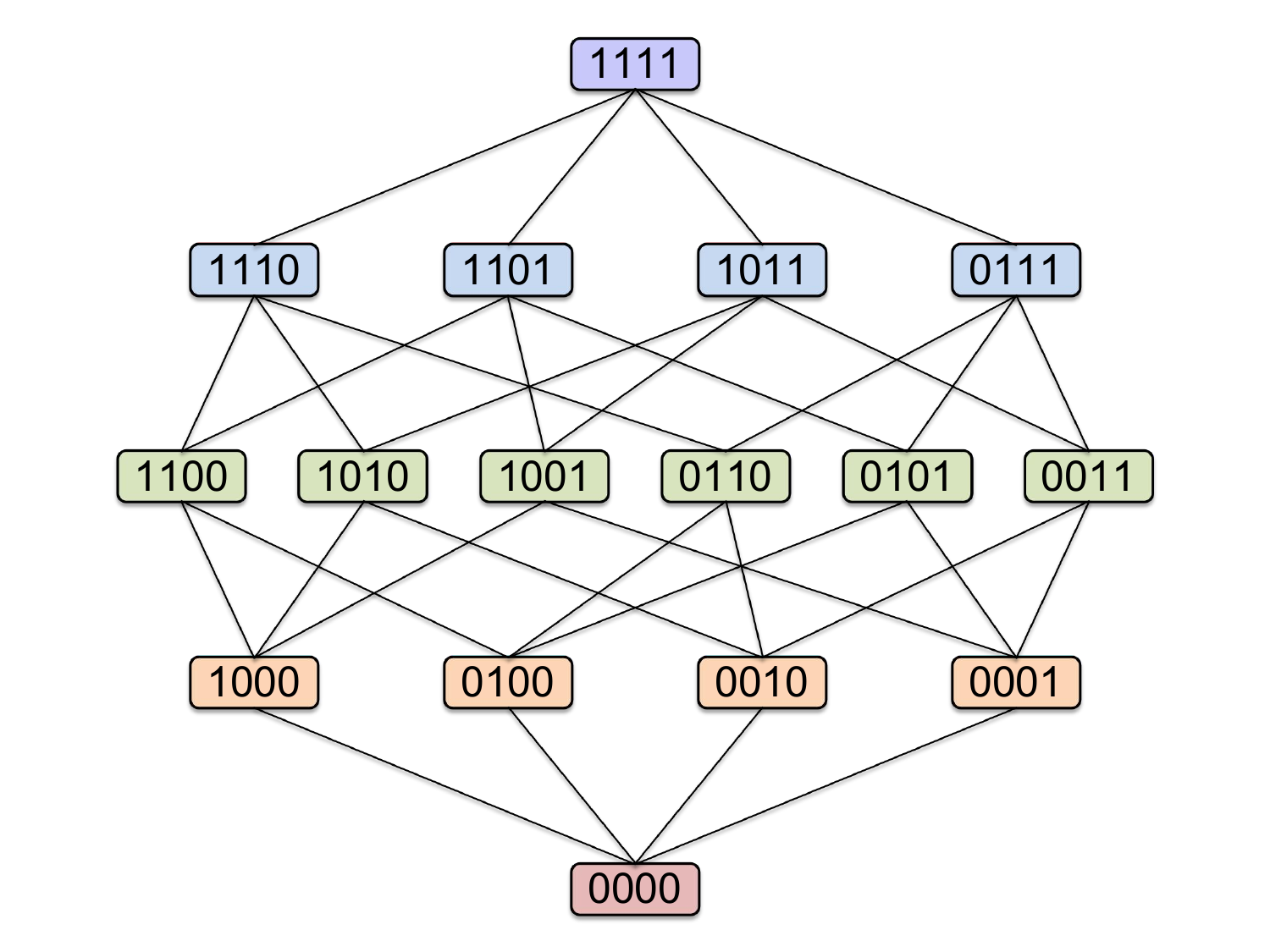}
    \caption{Schematic of the four-dimensional hypercube formed by the computational basis states of four qubits. Edges indicate nonzero Hamiltonian couplings between basis states, as defined by the Hamiltonian $H_\text{syn}=\sum_{k=1}^{N}\sigma_x^{(k)}$.}
    \label{fig:hypercube4sche}
\end{figure}

Fig.~\ref{fig:hypercube4sche} illustrates the 4-dimensional hypercube represented in the computational basis of 4 qubits, namely a Hilbert space with 16 states. The lines indicate the connectivity between vertices generated by the Hamiltonian. 
We use the perturbation theory to deal with the hypercube model. As mentioned in the main text, we assume independent and identical
single-qubit noise acting on each qubit. Using the first perturbation theory, namely Eq. (\ref{eq04}) and Eq. (\ref{hyperI}) for $N=4$, the steady state is
\vspace{0.5em}
\begin{equation}
\small
\ket{p^\mathrm{ss}}
= \frac{1}{16}
\begin{pmatrix}
{\color{hw0} 1 + 2(-p_{0\to1}+p_{1\to0})\,\csc^2\tau} \\   

{\color{hw1} 1 + (-p_{0\to1}+p_{1\to0})\,\csc^2\tau} \\   
{\color{hw1} 1 + (-p_{0\to1}+p_{1\to0})\,\csc^2\tau} \\   
{\color{hw2} 1} \\                                       
{\color{hw1} 1 + (-p_{0\to1}+p_{1\to0})\,\csc^2\tau} \\   
{\color{hw2} 1} \\                                       
{\color{hw2} 1} \\                                       
{\color{hw3} 1 + (p_{0\to1}-p_{1\to0})\,\csc^2\tau} \\   
{\color{hw1} 1 + (-p_{0\to1}+p_{1\to0})\,\csc^2\tau} \\   
{\color{hw2} 1} \\                                       
{\color{hw2} 1} \\                                       
{\color{hw3} 1 + (p_{0\to1}-p_{1\to0})\,\csc^2\tau} \\   
{\color{hw2} 1} \\                                       
{\color{hw3} 1 + (p_{0\to1}-p_{1\to0})\,\csc^2\tau} \\   
{\color{hw3} 1 + (p_{0\to1}-p_{1\to0})\,\csc^2\tau} \\   
{\color{hw4} 1 + 2(p_{0\to1}-p_{1\to0})\,\csc^2\tau}     
\end{pmatrix}
.
\label{perturIhyper4}
\end{equation}
Here the components of $\ket{p^{\mathrm{ss}}}$ are ordered according to the binary representation of the computational basis states. 
Specifically, the basis states are labeled as $\ket{b_1 b_2 b_3 b_4}$ with $b_i\in\{0,1\}$ and ordered by their binary index 
$b_1 b_2 b_3 b_4 = 0000, 0001, \ldots, 1111$, corresponding to the decimal indices $0$ through $15$. 
Thus, for example, $\ket{0000}$ is the first component of the vector and $\ket{1111}$ the last. Since we assume all qubits are identical and the hypercube is highly symmetric, the result depends only on the Hamming weight, i.e., the number of $\ket{1}$s, of the state. When the components are regrouped according to Hamming weight, the underlying symmetry of the hypercube becomes manifest, with the colors of the elements chosen to match those in Fig.~\ref{fig:hypercube4sche}.

\begin{figure*}[!htbp]
    \centering
    \begin{subfigure}[]{0.353 \textwidth}
        \centering
        \includegraphics[width=\textwidth]{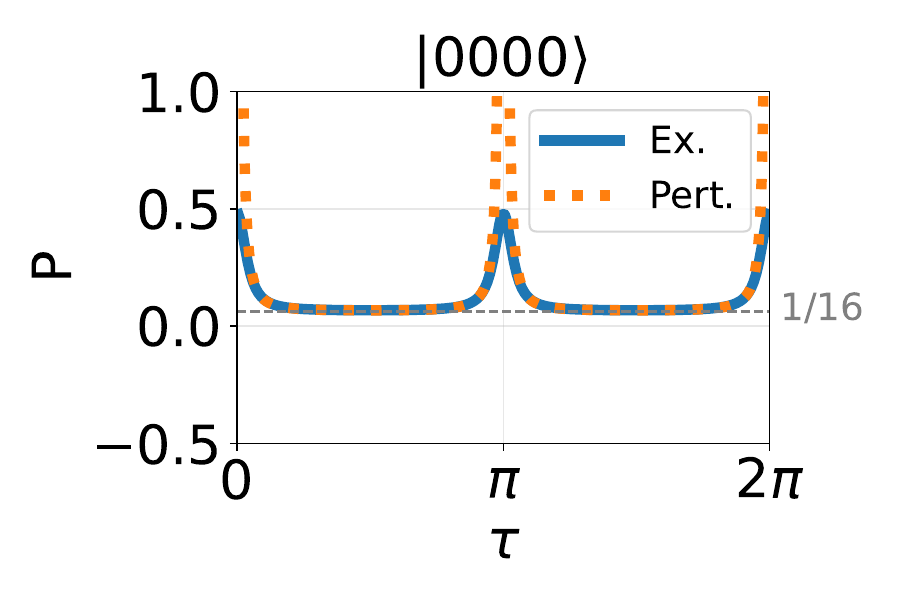}
    \end{subfigure}
    \begin{subfigure}[]{0.294 \textwidth}
        \centering
        \includegraphics[width=\textwidth]{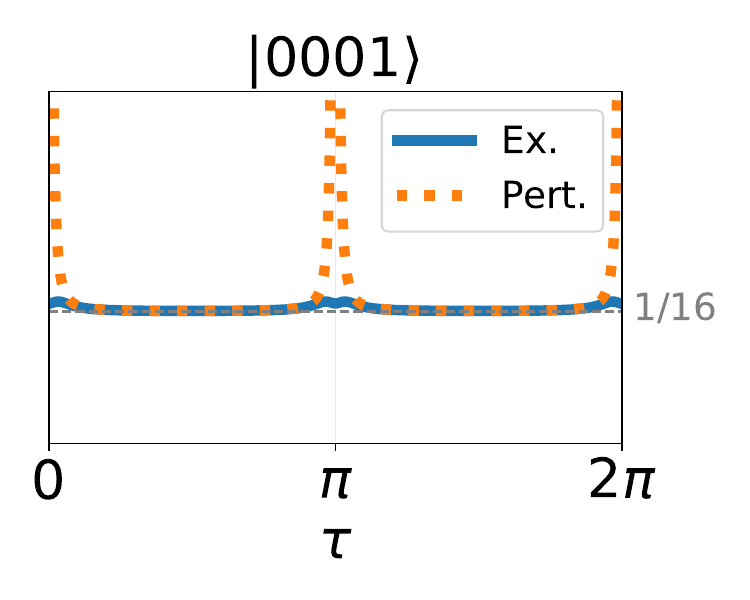}
    \end{subfigure}
    \begin{subfigure}[]{0.294 \textwidth}
        \centering
        \includegraphics[width=\textwidth]{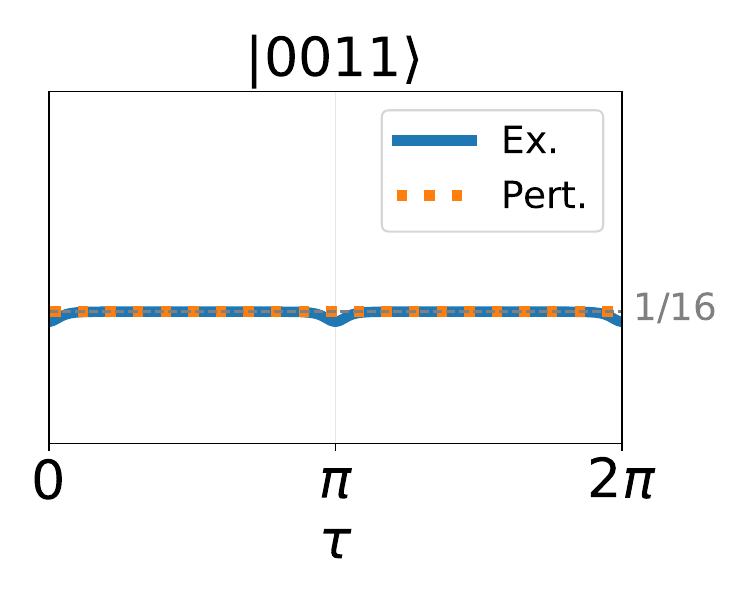}
    \end{subfigure}
    \begin{subfigure}[]{0.353 \textwidth}
        \centering
        \includegraphics[width=\textwidth]{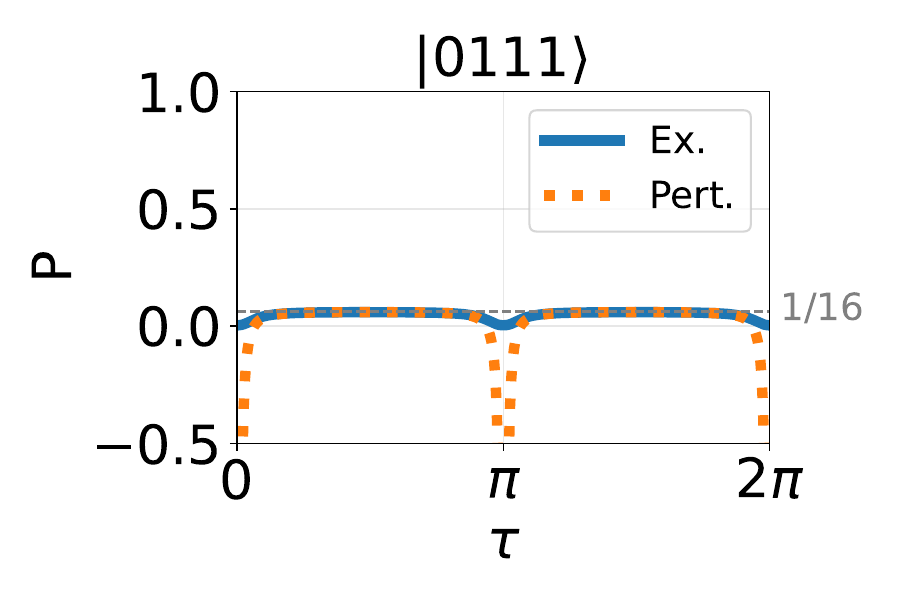}
    \end{subfigure}
    \begin{subfigure}[]{0.294 \textwidth}
        \centering
        \includegraphics[width=\textwidth]{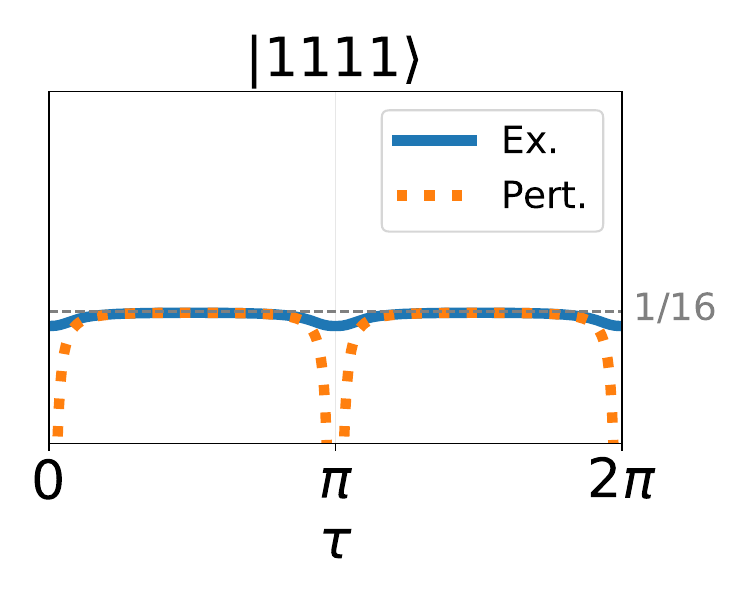}
    \end{subfigure}
    \caption{Steady-state probabilities on the 4-dimensional hypercube, grouped by the Hamming weight of the computational basis states. The perturbative approximations come from perturbation theory I (Eq.~(\ref{perturIhyper4})). Due to the hypercube's high symmetry and the assumption of uniform noise across qubits, the 16 basis states fall into five distinct classes, as indicated by the color grouping in Fig.~\ref{fig:hypercube4sche}. For example, the state $\ket{0001}$ represents the four equivalent single-excitation states $\{\ket{0001}, \ket{0010}, \ket{0100}, \ket{1000}\}$. The perturbative results (orange dashed lines) are compared with exact numerical solutions (blue solid lines) for error probabilities \( p_{1\to0} = 0.05 \) and \( p_{0\to1} = 0.01 \). Agreement is excellent except near \( \tau = 0, \pi, 2\pi \), where noise-induced effects become non-perturbative.}
    \label{fig:hypercube4overall}
\end{figure*}
 
\begin{figure*}[!htbp]
    \centering
    \begin{subfigure}[b]{0.32 \textwidth}
        \centering
        \includegraphics[width=\textwidth]{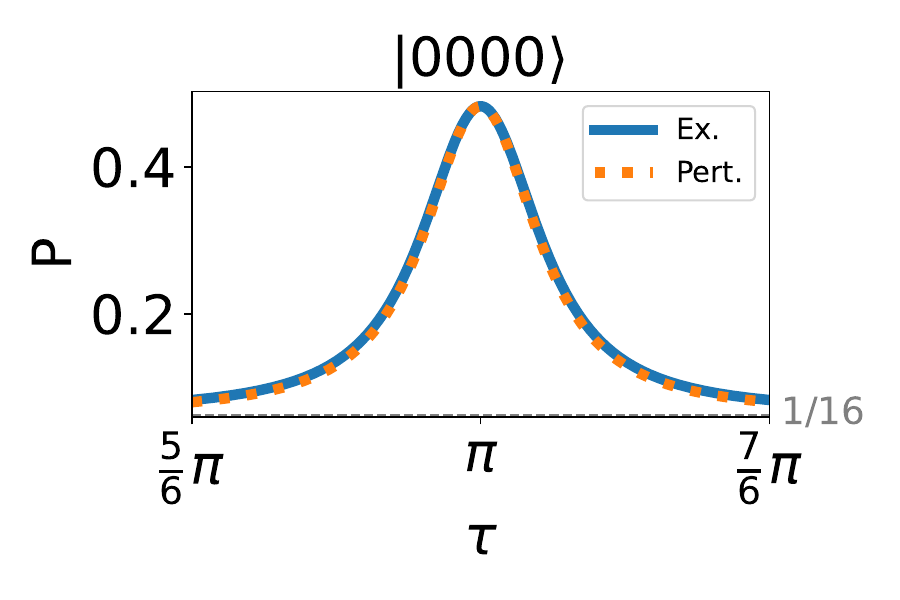}
    \end{subfigure}
    \begin{subfigure}[b]{0.32 \textwidth}
        \centering
        \includegraphics[width=\textwidth]{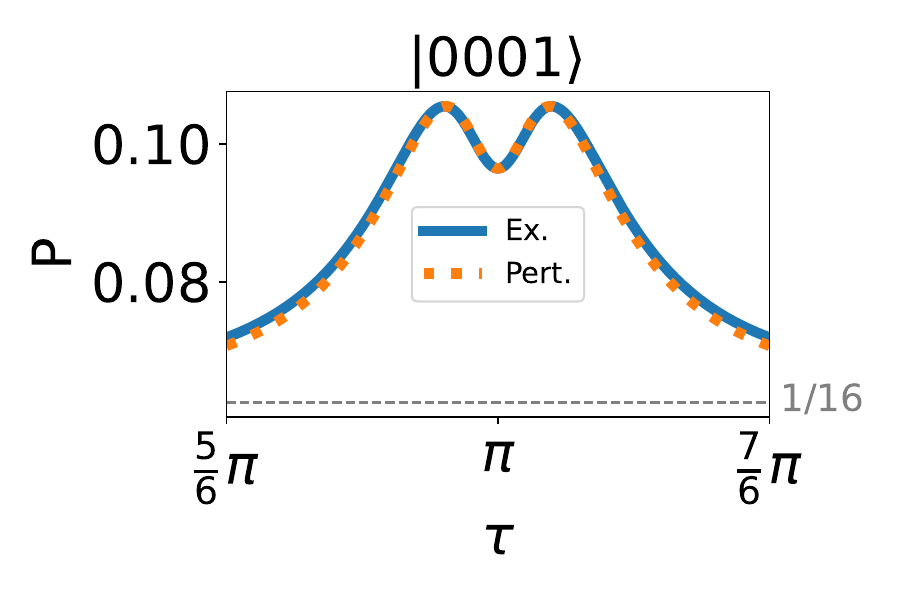}
    \end{subfigure}
    \begin{subfigure}[b]{0.32 \textwidth}
        \centering
        \includegraphics[width=\textwidth]{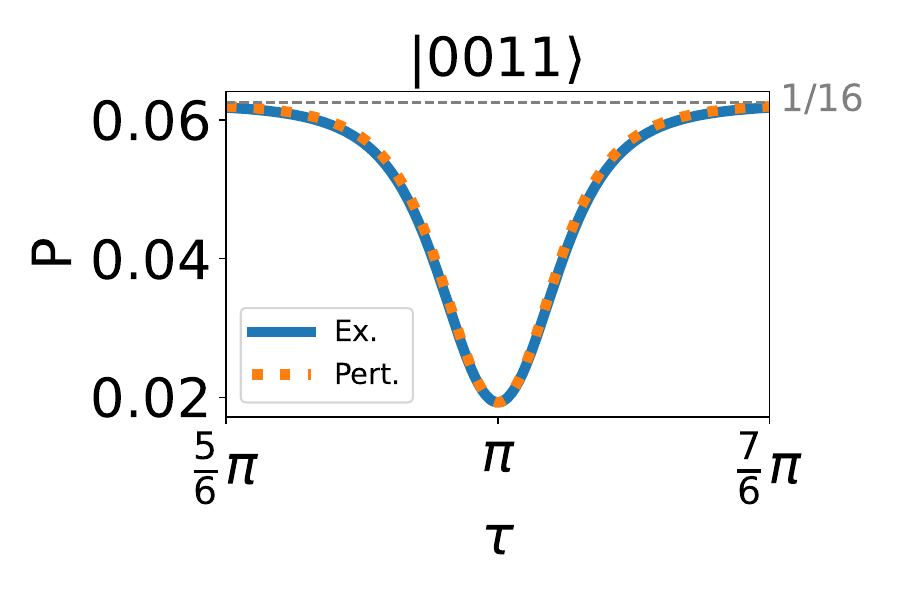}
    \end{subfigure}
    \begin{subfigure}[b]{0.32 \textwidth}
        \centering
        \includegraphics[width=\textwidth]{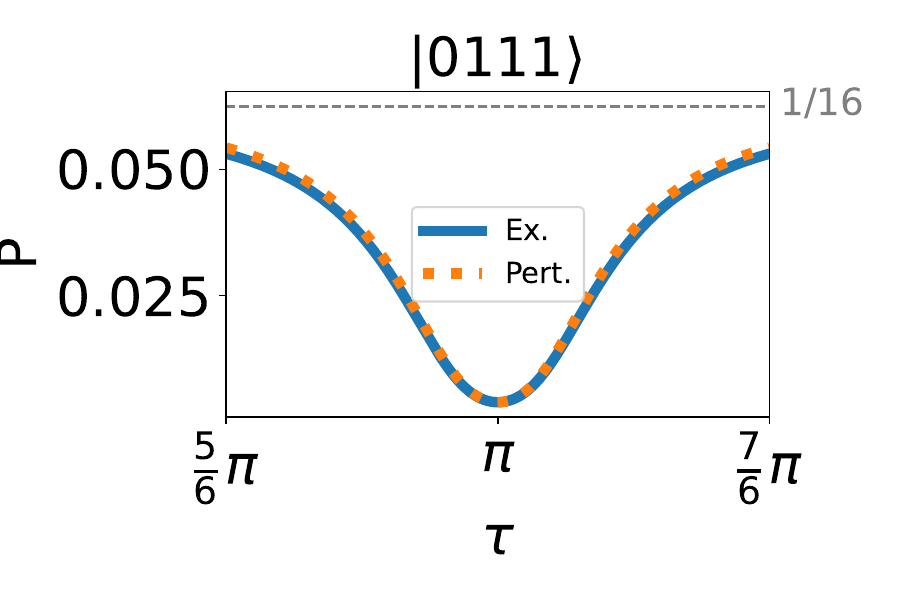}
    \end{subfigure}
    \begin{subfigure}[b]{0.32 \textwidth}
        \centering
        \includegraphics[width=\textwidth]{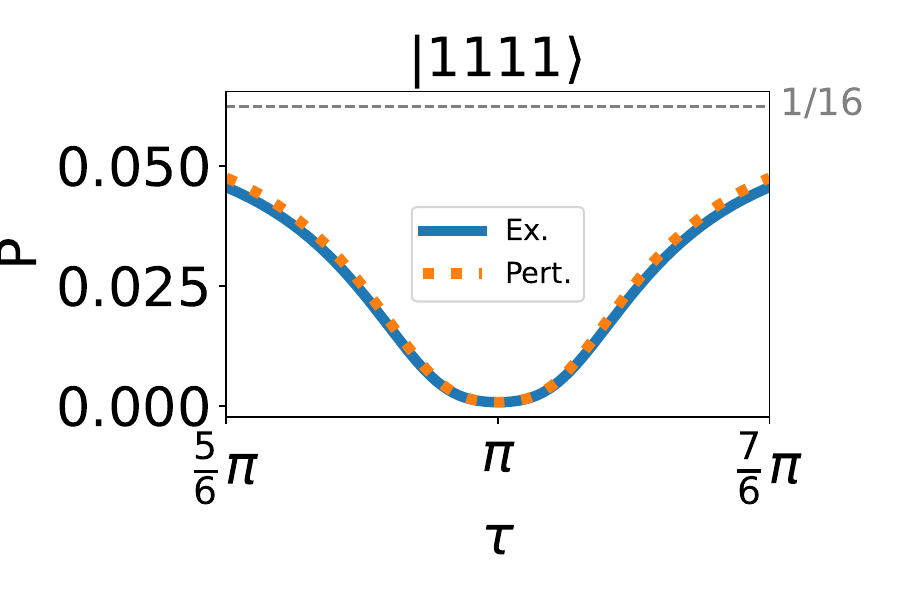}
    \end{subfigure}
    \caption{Comparison of the exact steady state with the perturbative approximation from perturbation theory II (Eq.~(\ref{hyperII})) for the four-qubit hypercube near the revival time $\tau_R=\pi$. The noise parameters are \(p_{1\to0} = 0.05\) and \(p_{0\to1} = 0.01\), assuming all the qubits are identical. We see that the physical ground state $\ket{0000}$ is the most populated at revival times, indicating low-temperature-like physics as explained in the text.}

    \label{fig:hypercube4ii}
\end{figure*}

Fig.~\ref{fig:hypercube4overall} shows the steady-state probabilities of representative states for the five distinct cases. We compare the exact solutions with the perturbative result Eq.~(\ref{perturIhyper4}).  Altogether, the hypercube has 16 basis states, and in the absence of noise, the steady-state probability is uniformly $1/16$, indicated by the dashed grey line. Away from the special sampling times $\tau = m\pi,\; m = 0,1,2,\ldots$, the perturbative expression agrees well with the exact numerical result, both remaining close to the noiseless value $1/16$. Recall that the mean recurrence time is fixed by the steady-state probability through $\langle n \rangle = 1/p$. For example, for the initial state $\ket{1111}$ and away from resonance, we have $p=1/16$, yielding $\langle n \rangle = 16$, consistent with the high-temperature limit predicted in the main text. In contrast, near $\tau = m\pi$, the perturbative result deviates significantly from the exact solution, and both depart strongly from the noiseless value. For the sampling intervals near revival time, Fig. \ref{fig:hypercube4ii} compares the exact result with the perturbative approximation near the revival time $\tau_R=\pi$ using Eq. (\ref{hyperII}).
By using perturbation theory II, the breakdown observed in the first perturbation theory is bypassed.

\section{Noise models}
\subsection{Generalized amplitude damping model}
Besides the Markov model used in the main text, we verify that a generalized amplitude damping (GAD) channel---capturing energy exchange with a finite-temperature environment---reproduces the same experimental trend \cite{Nielsen}. In each stroboscopic step of interval $\tau$, we apply a unitary $U(\tau)$ followed by a local GAD map with Kraus operators
\begin{equation}
\begin{aligned}
E_0&=\sqrt{p}\begin{pmatrix}1&0\\0&\sqrt{1-\gamma}\end{pmatrix},\qquad
E_1=\sqrt{p}\begin{pmatrix}0&\sqrt{\gamma}\\0&0\end{pmatrix},\\
E_2&=\sqrt{1-p}\begin{pmatrix}\sqrt{1-\gamma}&0\\0&1\end{pmatrix},\qquad
E_3=\sqrt{1-p}\begin{pmatrix}0&0\\\sqrt{\gamma}&0\end{pmatrix}.
\end{aligned}
\end{equation}
where $\gamma\in[0,1]$ is the damping strength and $p\in[0,1]$ sets the asymptotic thermal bias. Conditioning on no detection at the end of each stroboscopic step via the projector $Q$, the corresponding (unnormalized) post-measurement state reads
\begin{equation}
\rho(\tau)=
Q\Bigl(\sum_{k=0}^{3} E_k\,U(\tau)\rho_0 U^\dagger(\tau)\,E_k^\dagger\Bigr)Q,
\end{equation}
where the initial state is $\ket{\psi_0}$, i.e.,
\begin{equation}
\rho_0=\ket{\psi_0}\bra{\psi_0}.
\end{equation}
We take the target state to be the same, $\ket{\psi_0}$, so the detection projector is
\begin{equation}
P=\ket{\psi_0}\bra{\psi_0},
\qquad
Q=\mathcal{I}-P.
\end{equation}
The first-detection probability at the $n$th attempt is then
\begin{equation}
F_n=\mathrm{Tr}\!\left[
P\sum_{k=0}^{3} E_k\,U(\tau)\rho_{n-1} U^\dagger(\tau)\,E_k^\dagger
\right],
\qquad
\rho_{n}=
Q\left(\sum_{k=0}^{3} E_k\,U(\tau)\rho_{n-1} U^\dagger(\tau)\,E_k^\dagger\right)Q,
\label{KrausFn}
\end{equation}
and the mean recurrence time is
\begin{equation}
    \langle n \rangle = \sum_{n=1}^\infty nF_n.
\end{equation}

\begin{figure}
    \centering
    \includegraphics[width=0.5\linewidth]{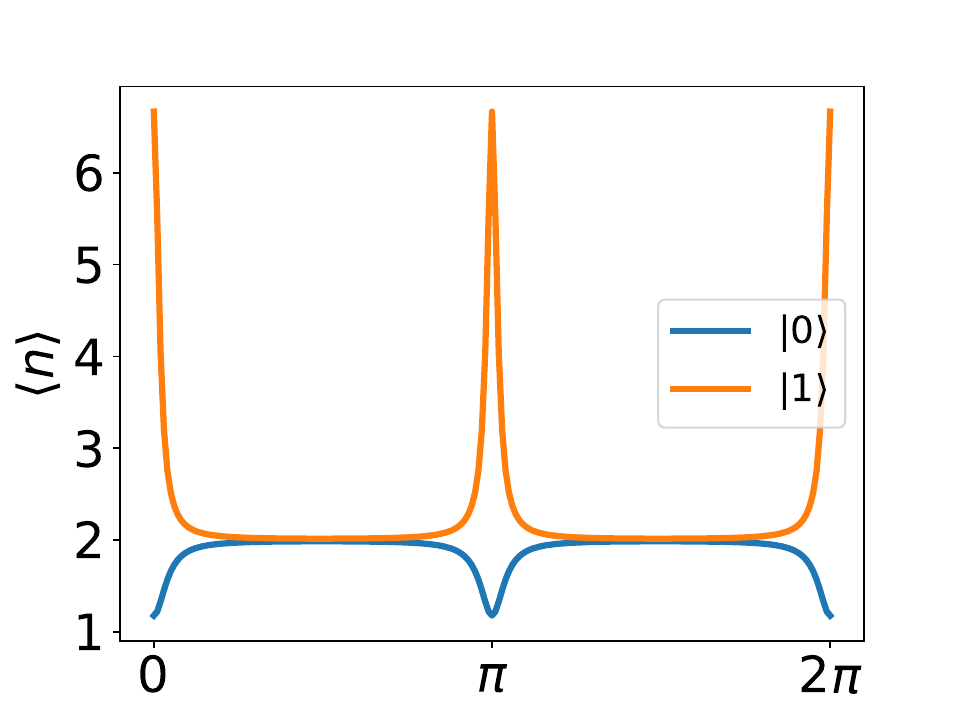}
    \caption{Mean recurrence time $\langle n\rangle$ for a single-qubit generalized amplitude damping (GAD) model with $\gamma=0.02$ and $p=0.85$, plotted as a function of the stroboscopic interval $\tau$. The GAD prediction exhibits the same qualitative $\tau$-dependence as the experimental trend reported in the main text.}
    \label{fig:GAD}
\end{figure}

We extend the above construction to a two-qubit system subject to independent local noise on each qubit. In each stroboscopic step of duration $\tau$, we apply a unitary $U(\tau)$ acting on the two-qubit Hilbert space, followed by local generalized amplitude damping channels acting separately on qubits $A$ and $B$. Let ${E_k^{(A)}}$ and ${E_l^{(B)}}$ denote the single-qubit GAD Kraus operators for qubits $A$ and $B$, respectively, with possibly different parameters $(\gamma_A,p_A)$ and $(\gamma_B,p_B)$. The two-qubit channel is then given by the tensor-product Kraus set
\begin{equation}
\mathcal{E}(\rho)=
\sum_{k,\ell=0}^{3}
\left(E_k^{(A)}\otimes E_\ell^{(B)}\right)
\rho
\left(E_k^{(A)}\otimes E_\ell^{(B)}\right)^\dagger,
\end{equation}
which contains $16$ Kraus operators. This construction describes independent energy exchange of each qubit with its own finite-temperature environment.

\begin{figure}
    \centering
    \includegraphics[width=0.8\linewidth]{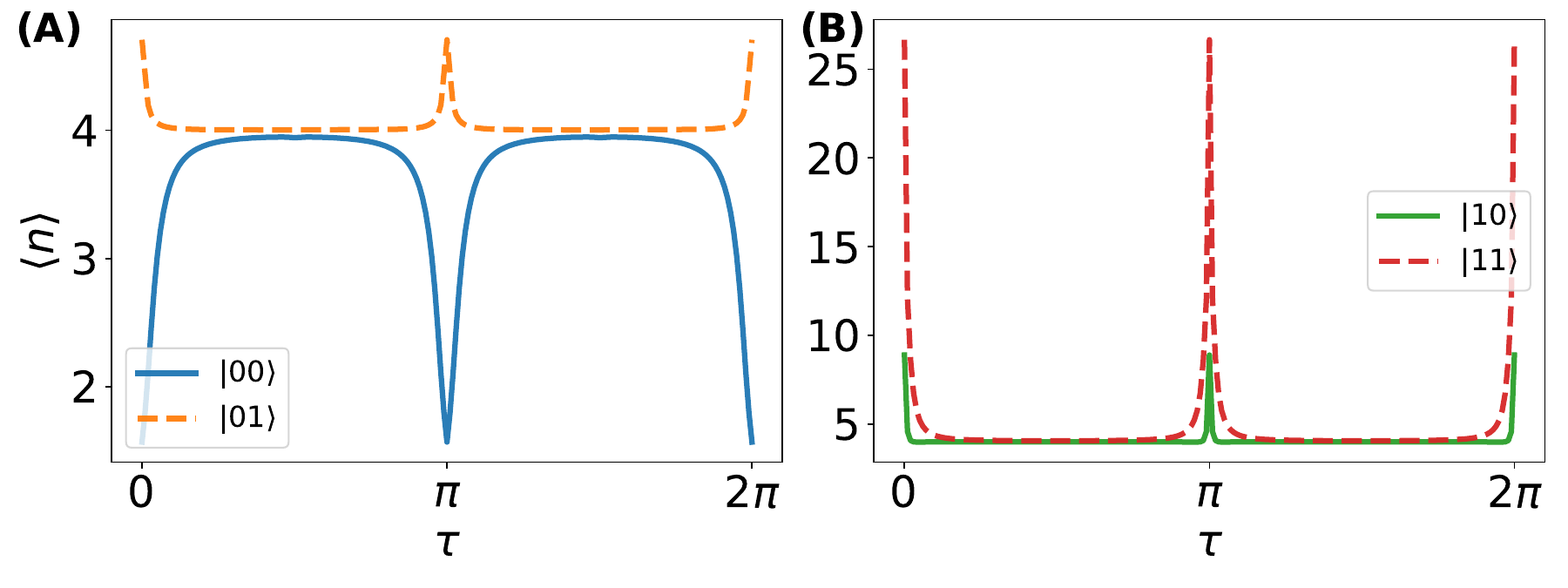}
    \caption{Mean recurrence time $\langle n\rangle$ for a two-qubit generalized amplitude damping (GAD) model with $\gamma_A=0.02$, $p_A=0.85$, $\gamma_B=0.03$, $p_B=0.75$.}
    \label{fig:GAD2}
\end{figure}

Fig. \ref{fig:GAD} and Fig. \ref{fig:GAD2} shows the $\langle n \rangle$ using the GAD model. While the GAD channel can thus also account for the observed behavior, the Markov model adopted in the main text is more directly aligned with our effective description of the device noise and, importantly, admits a transparent analytical treatment of the stroboscopic dynamics; by Occam's razor, we therefore favor it as the most economical framework for modeling.

\subsection{Other noise models}
Other noise models, however, fail to describe the experimental behavior. 
In particular, unital noise models such as depolarization and bit-phase flip channels do not reproduce the observed $\tau$-dependent resonance structure. 
For the single-qubit depolarizing channel we use the Kraus representation
\begin{equation}
E_0=\sqrt{1-\eta}\,\mathcal{I},\quad
E_1=\sqrt{\eta/3}\,\sigma_x,\quad
E_2=\sqrt{\eta/3}\,\sigma_y,\quad
E_3=\sqrt{\eta/3}\,\sigma_z,
\end{equation}
where $\eta\in[0,1]$ is the depolarizing strength. 
Likewise, the bit-phase flip channel is
\begin{equation}
E_0=\sqrt{1-q}\,\mathcal{I},\quad
E_1=\sqrt{q}\,\sigma_y,
\end{equation}
with $q\in[0,1]$. 
Using Eq.~(\ref{KrausFn}), now with a general set of Kraus operators describing the noise process
(for multi-qubit systems given by tensor products of the single-qubit Kraus operators), we find that the first-recurrence dynamics yield a $\tau$-independent mean recurrence time that saturates to a constant equal to the Hilbert-space dimension \cite{Sinkovicz2015},
\begin{equation}
\langle n\rangle \simeq d,
\end{equation}
in stark contrast with the strong $\tau$-dependent peaks and dips observed experimentally.

The zero-temperature amplitude damping (AD) model fails for a different reason: it neglects thermal excitation from $\ket{0}$ to $\ket{1}$. 
Its Kraus operators are
\begin{equation}
E_0=
\begin{pmatrix}
1 & 0\\
0 & \sqrt{1-\gamma}
\end{pmatrix},
\quad
E_1=
\begin{pmatrix}
0 & \sqrt{\gamma}\\
0 & 0
\end{pmatrix},
\end{equation}
where $\gamma\in[0,1]$ is the damping strength. 
As a consequence, at a revival the recurrence to $\ket{0}$ is essentially immediate, giving $\langle n\rangle \approx 1$, whereas recurrence to $\ket{1}$ becomes effectively impossible in the long-time limit, implying $\langle n\rangle\to\infty$. 
This pathology is avoided by the finite-temperature GAD channel used above, which incorporates both relaxation and excitation processes through the thermal-bias parameter $p$.

\section{Derivation of the Perturbation theories}
\subsection{Perturbation theory I: away from revival times}
As mentioned in the main text, the unperturbed steady state satisfies
\begin{equation}
    \ket{p^{\mathrm{ss}}}_0 =\frac{1}{2^N}\,(1,1,\ldots,1)^{\mathsf{T}}
    \label{pss0}
\end{equation}
where the subscript 0 denotes the unperturbed solution, and $N$ is the number of qubits. 
This steady state holds true and is unique if the system is ergodic. Our objective is to compute the stationary distribution \( \ket{p^{\mathrm{ss}}} \) perturbatively in the weak noise limit. Specifically, we expand the steady state as
\begin{equation}
    \ket{p^{\mathrm{ss}}} = \ket{p^{\mathrm{ss}}}_0 + \epsilon \ket{p^{\mathrm{ss}}}_1 + \mathcal{O}(\epsilon^2),
    \label{PerturbativeExpansion}
\end{equation}
where \( \ket{p^{\mathrm{ss}}}_1 \) is the first-order correction. Since $G \ket{p^\mathrm{ss}}_0 = \ket{p^\mathrm{ss}}_0$, \( G - \mathcal{I} \) is singular and non-invertible. This fact will be important when we construct the perturbation theory.

For the steady state of the perturbed system, we write
\begin{align}
    G'G \ket{p^\mathrm{ss}} &= \ket{p^\mathrm{ss}}, \\
    (\mathcal{I}+\epsilon C)G\left(\ket{p^\mathrm{ss}}_0+\epsilon\ket{p^\mathrm{ss}}_1+\dots\right) &= \ket{p^\mathrm{ss}}_0+\epsilon\ket{p^\mathrm{ss}}_1+\dots
\end{align}
At first order in \( \epsilon \), we have
\begin{equation}
CG\ket{p^\mathrm{ss}}_0 + G\ket{p^\mathrm{ss}}_1 = \ket{p^\mathrm{ss}}_1,
\end{equation}
which leads to
\begin{equation}
(\mathcal{I} - G)\ket{p^\mathrm{ss}}_1 = C\ket{p^\mathrm{ss}}_0.
\label{I-G}
\end{equation}
However, since \( \mathcal{I} - G \) is non-invertible, we use a projection method to solve the problem perturbatively. Normalization of \( \ket{p^\mathrm{ss}} \) implies
\begin{equation}
\sum_{i=0}^{2^N-1} \langle i | p^\mathrm{ss} \rangle_1 = 0,
\end{equation}
or equivalently,
\begin{equation}
J_N \ket{p^\mathrm{ss}}_1 = 0,
\label{JN}
\end{equation}
where \( J_N \) is the all-ones matrix. We employ Eq. (\ref{JN}) in the perturbation theory, as we now show. Thus, we regularize the equation:
\begin{equation}
\left(\mathcal{I} - G + \alpha J_N\right) \ket{p^\mathrm{ss}}_1 = C \ket{p^\mathrm{ss}}_0,
\end{equation}
where $\alpha$ is a non-zero constant. We can choose $\alpha=1 /2^N$ for simplicity, though any non-zero value of $\alpha$ would work. For generic cases, the matrix \( O = \mathcal{I} - G + \frac{1}{2^N} J_N \) is invertible, allowing us to write
\begin{equation}
\ket{p^\mathrm{ss}}_1 = O^{-1} C \ket{p^\mathrm{ss}}_0.
\label{firstorderI}
\end{equation}
The key step is to introduce a suitable regularization. Note that this perturbation theory fails when $G = \mathcal{I}$ and $G^2 = \mathcal{I}$, where the operator $O$ becomes singular. In such situations, we must employ another perturbation theory to properly describe the steady-state corrections. In general, the case $G^2=\mathcal{I}$ admits a perturbative expansion of the same form, whereas this is not true for $G=\mathcal{I}$. As a special case, the $G=\mathcal{I}$ scenario is discussed separately in the next section.

\subsection{Perturbation theory II: near revival times}
Near these revival times, the perturbative correction from perturbation theory I diverges and the approach breaks down. To capture the dynamics in this regime, we introduce a second small parameter quantifying the deviation from a revival time, and develop a new perturbative framework that treats both the noise strength and proximity to $G(\tau) = \mathcal{I}$ as small. For $\tau=\tau_R+\delta\tau$, where $G(\tau_R)=\mathcal{I}$,
\begin{equation}
    G_{ij} = \left|\left\langle i \middle| e^{-iH\tau_R - iH(\delta\tau)} \middle| j \right\rangle\right|^2 
    \simeq \left|\left\langle i \middle| \left(1 - iH(\delta\tau) - \frac{H^2(\delta\tau)^2}{2} \right) \middle| j \right\rangle\right|^2.
\end{equation}
This corresponds to a near-revival expansion around a stroboscopic identity of the Floquet evolution operator \cite{Bukov}. Therefore, we get the expression for $G_{ij}$
\begin{equation}
    G_{ij}\simeq|\delta_{ij}-iH_{ij}\delta\tau-\frac{1}{2}(H^2)_{ij}(\delta\tau)^2|^2.
\end{equation}
When $i=j$,
 \begin{equation}
    G_{ii} \simeq |1-iH_{ii}\delta\tau-\frac{1}{2}(H^2)_{ii}(\delta\tau)^2|^2
    \simeq 1+((H_{ii})^2-(H^2)_{ii})(\delta \tau)^2.
\end{equation}
And when $i\neq j$
\begin{equation}
    G_{ij}\simeq|-iH_{ij}\delta\tau-\frac{1}{2}(H^2)_{ij}(\delta\tau)^2|^2\simeq(H_{ij})^2(\delta\tau)^2.
\end{equation}
Rewrite the diagonal terms of the square of the Hamiltonian using $\sum_k \ket{k}\bra{k}=\mathcal{I}$
\begin{equation}
    (H^2)_{ii}=\sum_k \langle i|H|k\rangle\langle k|H|i\rangle =\sum_k|H_{ik}|^2,
\end{equation}
so the diagonal terms of operator $G$ can be further written as
\begin{equation}
    G_{ii} \simeq 1-\sum_{k\neq i}|H_{ik}|^2(\delta \tau)^2.
\end{equation}
Hence,
\begin{equation}
    G\simeq\mathcal{I}-(\delta \tau)^2W
\end{equation}
where the elements of $W$ are
\begin{equation}
W_{ij} =
\begin{cases}
- |H_{ij}|^2, & i \neq j \\
\sum_{k \neq i} |H_{ik}|^2, & i=j
\end{cases}
\label{W}
\end{equation}
Now we assume $G' = \mathcal{I} + \epsilon C$, where $|\epsilon| \ll 1$. 
The steady-state condition $G'G\ket{p^\mathrm{ss}} = \ket{p^\mathrm{ss}}$ then reads
\begin{equation}
    (\mathcal{I} + \epsilon C)(\mathcal{I} - (\delta\tau)^2 W)\ket{p^\mathrm{ss}} = \ket{p^\mathrm{ss}}.
\end{equation}
Assuming $\epsilon$ and $(\delta\tau)^2$ are of the same order, and neglecting higher-order terms 
$\mathcal{O}(\epsilon^2)$, we find
\begin{equation}
    [\epsilon C - (\delta\tau)^2 W]\ket{p^\mathrm{ss}} + \ket{p^\mathrm{ss}} = \ket{p^\mathrm{ss}}.
\end{equation}
Subtracting $\ket{p^\mathrm{ss}}$ from both sides yields
\begin{equation}
    [\epsilon C - (\delta\tau)^2 W]\ket{p^\mathrm{ss}} = 0,
    \label{PerturbII}
\end{equation}
so that $\ket{p^\mathrm{ss}}$ is obtained as the normalized null vector of 
$\epsilon C - (\delta\tau)^2 W$. It works in the vicinity of revivals, where the influence of weak noise is most pronounced.

\section{Details of the quantum processor}
\begin{table}[ht]
\centering
\begin{tabular}{lccc}
\toprule
Parameter & Qubit 28 & Qubit 130 & Qubit 132 \\
\midrule
$T_1$ ($\mu$s) & 196.25 & 187.04 & 172.24 \\
$T_2$ ($\mu$s) & 205.34 & 171.04 & 225.32 \\
\addlinespace
Readout assignment error & $8.301\times10^{-3}$ & $5.615\times10^{-3}$ & $5.981\times10^{-3}$ \\
$P(0|1)$ & 0.0144 & 0.00708 & 0.01001 \\
$P(1|0)$ & 0.0022 & 0.00415 & 0.00195 \\
Readout length (ns) & 1560 & 1560 & 1560 \\
\addlinespace
Single-qubit gate length (ns) & 24 & 24 & 24 \\
ID gate error & $2.586\times10^{-4}$ & $4.465\times10^{-4}$ & $1.87\times10^{-4}$ \\
RX gate error & $2.586\times10^{-4}$ & $4.465\times10^{-4}$ & $1.87\times10^{-4}$ \\
SX gate error & $2.586\times10^{-4}$ & $4.465\times10^{-4}$ & $1.87\times10^{-4}$ \\
X gate error & $2.586\times10^{-4}$ & $4.465\times10^{-4}$ & $1.87\times10^{-4}$ \\
$R_Z$ gate error & 0 & 0 & 0 \\
\bottomrule
\end{tabular}
\caption{Calibration parameters of physical qubits 28, 130, and 132 on IBM Quantum processor ibm\_fez.}
\label{tab:calib_ibm_fez}
\end{table}
The experiments were performed on the IBM Quantum Heron-class superconducting quantum processor ibm\_fez. We used physical qubits 28, 130 and 132 of the device. The relevant calibration parameters of the qubits, including coherence times, readout errors, and single-qubit gate fidelities, are summarized in Table~\ref{tab:calib_ibm_fez}. All circuits were executed using IBM Quantum Runtime.

\subsection*{Comparison of fitted noise parameters with device calibration data}

The effective noise parameters $p_{1\to0}$ and $p_{0\to1}$ entering the Markov model are obtained from a fit to the measured recurrence times. It is instructive to compare these fitted values with independently reported calibration quantities of the IBM quantum processor \texttt{ibm\_fez} (Table~1), which characterize $T_1$ relaxation and readout assignment errors for each qubit.

Each stroboscopic cycle consists of a single-qubit $R_x$ gate (24\,ns) followed by a projective measurement (1560\,ns). Two hardware-level mechanisms contribute to effective bit-flip errors per cycle: (i) $T_1$ relaxation, predominantly during the readout interval, and (ii) anisotropic readout assignment errors. The readout is strongly biased toward reporting $|0\rangle$: for all three qubits the conditional misassignment probability satisfies $P(0|1)\gg P(1|0)$. In the present device this bias is consistent with both $T_1$ decay during the readout window and asymmetries of the calibrated readout channel.

For the single-qubit experiment (qubit~28), the $T_1$-based relaxation probability per cycle is
\[
p_{1\to0}^{(T_1)} \approx \frac{t_{\mathrm{readout}}}{T_1}
= \frac{1560}{196\,250} \approx 0.0079,
\]
in good agreement with the fitted value $p_{1\to0}=0.0086$. The fitted excitation probability $p_{0\to1}=0.0012$ is of the same order as the calibrated misassignment probability $P(1|0)=0.0022$. 

For the two-qubit experiment (qubits~130 and~132), the $T_1$-based per-qubit decay probabilities are $p_{1\to0}^{(T_1)}\approx0.0085$ (qubit~130) and $\approx0.0092$ (qubit~132), while the fitted values $p_{1\to0,a}=0.0189$ and $p_{1\to0,b}=0.0188$ are larger by roughly a factor of two. This discrepancy indicates additional circuit-level noise contributions in the two-qubit protocol, such as idle-time relaxation, crosstalk, or synchronization overhead, which are absorbed into the effective Markov parameters.

Despite these differences in absolute magnitude, the \emph{asymmetry ratio} $\frac{p_{0\to1}}{p_{1\to0}}$
is well reproduced. For the single-qubit experiment the fitted ratio is $\frac{0.0012}{0.0086}=0.14$,
in close agreement with the calibrated readout anisotropy
\[
\frac{P(1|0)}{P(0|1)}=\frac{0.0022}{0.0144}=0.15.
\]
This shows that the directional bias of the effective noise—the property that controls the steady-state asymmetry and the peak/dip structure near resonance—is captured correctly by the calibrated hardware asymmetry. 

For the two-qubit case, the fitted ratios ($0.14$ and $0.12$) are more uniform than the qubit-resolved readout anisotropies ($0.59$ for qubit~130 and $0.19$ for qubit~132), again indicating that additional asymmetric noise sources partially average over the bare readout structure.

Genuine thermal excitation at dilution-refrigerator temperatures is negligible: for a typical transmon frequency $\omega\approx5$\,GHz, the Boltzmann factor $\exp(-\hbar\omega/k_B T)$ is vanishingly small. The effective ratio $p_{0\to1}/p_{1\to0}\approx0.14$ should therefore not be interpreted as a true Boltzmann factor but rather as an operational parameter of the effective Markov model, set by the combined asymmetry of relaxation and readout.

In summary, the Markov noise model is an effective description in which the two parameters $p_{1\to0}$ and $p_{0\to1}$ per qubit absorb multiple hardware-level noise mechanisms—$T_1$ relaxation, anisotropic readout assignment, crosstalk, and circuit-level overhead—into a minimal representation. Their absolute values cannot be inferred from single calibration numbers alone, but their scale and, more importantly, their asymmetry are consistent with the independently measured device characteristics.